\RequirePackage{fix-cm}

\documentclass[smallextended,nospthms]{svjour3}

\usepackage{graphicx}
\usepackage{xcolor}
\usepackage{fontawesome5}
\usepackage{multicol}
\usepackage{tabularx}
\usepackage{booktabs}
\usepackage{url}
\usepackage{caption}
\usepackage{tcolorbox}
\usepackage{colortbl}
\usepackage{enumitem}
\usepackage{hyphenat}
\usepackage{makecell,rotating}
\usepackage{multirow}
\usepackage{float}
\usepackage{soul}
\usepackage{xspace}
\usepackage{amsmath}
\usepackage{amssymb}
\usepackage{bm}
\usepackage[]{mdframed}
\usepackage{longtable}
\usepackage{subcaption}
\usepackage{hyperref}

\newboolean{showcomments}
\setboolean{showcomments}{false}
\newboolean{peerreview}
\setboolean{peerreview}{false}

\definecolor{gray75}{gray}{.25}
\definecolor{gray50}{gray}{.5}
\definecolor{gray40}{gray}{.6}
\definecolor{gray30}{gray}{.7}
\definecolor{gray25}{gray}{.75}
\definecolor{gray20}{gray}{.8}
\definecolor{gray15}{gray}{.85}
\definecolor{gray10}{gray}{.9}
\definecolor{gray05}{gray}{.95}
\definecolor{darkgreen}{HTML}{317f43}
\definecolor{redbg}{HTML}{F2968F}
\definecolor{greenbg}{HTML}{CDE4AE}
\definecolor{goalcolor}{HTML}{fffff2}
\definecolor{rqboxcolor}{HTML}{f2f2ff}
\definecolor{rqanswercolor}{HTML}{faf9f5}
\definecolor{takeawaycolor}{HTML}{f2fff2}

\newcommand{\rqbox}[2]{%
	\medskip%
	\noindent\fcolorbox{black}{gray05}{%
		\parbox{0.985\linewidth}{%
			\textbf{\faSearch\ RQ$_{#1}$.} \textit{#2} %
		}%
	}%
	\medskip%
}%

\newcommand{\rqanswer}[2]
{
    \smallskip
    \begin{mdframed}[style=boxstyle5]
    {\textbf{\faHandPointRight\ Answer to RQ$_{#1}$.} #2}
    \end{mdframed}
}

\mdfdefinestyle{boxstyle4}{%
    linecolor=black,linewidth=2pt,%
    leftmargin=0cm,rightmargin=0cm,%
    roundcorner=5pt,
    topline=true,bottomline=true,rightline=false,leftline=false,%
    backgroundcolor=black!10,
    frametitlebackgroundcolor=black
}
\mdfdefinestyle{boxstyle5}{%
    linecolor=darkgreen,
    linewidth=3pt,%
    leftmargin=0cm,rightmargin=0cm,%
    roundcorner=5pt,
    topline=false,bottomline=false,rightline=true,leftline=true,%
    backgroundcolor=darkgreen!10
}
\newcommand{\prompt}[2]
{
    \medskip
    \begin{mdframed}[style=boxstyle4]
    {\underline{\textbf{#1}}\\#2}
    \end{mdframed}
    \smallskip
}

\newcommand\ques[2][]{$Q_{#2}^{#1}$}

\newcommand\rqq[1]{\textbf{RQ$_{#1}$}}

\definecolor{DarkOrchid}{rgb}{0.6, 0.2, 0.8}

\ifthenelse{\boolean{showcomments}}
{

}
{
}

\newcommand\claude{\textsc{Claude 3.7 Sonnet}\xspace}
\newcommand\claudeShort{\textsc{Claude}\xspace}
\newcommand\abbrevclaude{$C$\xspace}

\newcommand\gpt{\textsc{GPT-4o}\xspace}
\newcommand\gptShort{\textsc{GPT}\xspace}
\newcommand\abbrevgpt{$G$\xspace}

\newcommand\deepseekLong{\textsc{DeepSeek V4 Flash} 0731\xspace}
\newcommand\deepseekShort{\textsc{DeepSeek V4 Flash}\xspace}

\newcommand\llmForFirstPatch{\textit{LLMFor1stPatch}\xspace}
\newcommand\selectedIsHuman{\textit{SelectedPatchIsHuman}\xspace}

\newcommand\acaAllParticipants{111\xspace}
\newcommand\acaValidParticipants{79\xspace}
\newcommand\indParticipants{14\xspace}
\newcommand\allValidParticipants{93\xspace}
\newcommand\acaObservations{316\xspace}
\newcommand\indObservations{56\xspace}
\newcommand\allObservations{372\xspace}

\newcommand\answersYesAck{101\xspace}
\newcommand\answersNoAck{271\xspace}

\newcommand\codetheme[1]{\textit{``#1''}\xspace}
\newcommand\answerVerbatim[1]{\textcolor{black!60}{\textit{`#1'}}\xspace}

\begin{document}

\title{Human Reviewers' Attitudes Toward LLM-generated Security Patches: A Controlled Experiment}

\titlerunning{Human Reviewers' Attitudes Toward LLM-generated Security Patches}

\author{Emanuele Iannone \and Maria Camporese \and Nicolás E. Díaz Ferreyra \and G\'abor Antal \and P\'eter Heged\H{u}s \and Fabio Massacci \and Riccardo Scandariato}

\institute{Emanuele Iannone, Nicolás E. Díaz Ferreyra, Riccardo Scandariato \at
  Institute of Software Security, Hamburg University of Technology, Hamburg, Germany \\
  \email{emanuele.iannone@tuhh.de}, 
  \email{nicolas.diaz-ferreyra@tuhh.de},  
  \email{riccardo.scandariato@tuhh.de},
  \and
  G\'abor Antal, P\'eter Heged\H{u}s \at
  University of Szeged, Szeged, Hungary and FrontEndART Ltd., Szeged, Hungary\\
  \email{antal@inf.u-szeged.hu},
  \email{hpeter@inf.u-szeged.hu},
  \and
  Maria Camporese \at
  University of Trento, Trento, Italy \\
  \email{maria.camporese@unitn.it},
  \and
  Fabio Massacci \at
  University of Trento, Trento, Italy and Vrije Universiteit Amsterdam, Amsterdam, The Netherlands \\
  \email{fabio.massacci@ieee.org}
}

\authorrunning{Iannone et al.}

\date{Received: date / Accepted: date}

\maketitle

\begin{abstract}
Existing research in Automated Vulnerability Repair (AVR) has predominantly focused on developing techniques that produce more effective patches.
With the rise of generative AI (GenAI) models powering AVR tools, understanding how humans perceive AI-generated security patches becomes relevant, particularly how useful and trustworthy they are considered.

This paper presents a controlled experiment involving \acaAllParticipants Master's students from three universities and \indParticipants professionals from three companies working in software development and cybersecurity.
The experiment highlights the factors and motivations that influence human acceptance of AI-generated security patches.
Participants analyzed four real-world vulnerabilities, evaluated two plausible patches for each, selected the one they believed to be the most adequate fix, and explained their choice.
The two patches were generated by two LLMs, namely \gpt and \claude, though the participants were not aware of their involvement.
After each selection, participants were told that one of the two originated from a human, and given the opportunity to reconsider their choice and to motivate it.

The results show that, beyond correctness, participants favor patches that clearly explain how the vulnerability was fixed, preserved the original functionality, and appeared easy to install. 
There were also signals of a \textit{provenance-based bias}: a tendency to switch from the initially selected patch to the other attributed to a human, even though all patches originated from GenAI.
This bias was even explicitly admitted by some participants.
Participants who acknowledged patch provenance were also more likely to switch patches; nevertheless, most neglected this information altogether. 
The paper concludes with recommendations for AVR researchers and tool developers on how to leverage these findings.
\end{abstract}


\section{Introduction}
\label{sec:intro}

Automated Vulnerability Repair (AVR) is the task of fixing vulnerable code (e.g., a function) by automatically generating \textit{security patches}~\cite{hu:usenix2025:avr:sok}.
Research on AVR has predominantly focused on developing novel techniques to produce more accurate and effective candidate patches.
The typical way to assess the effectiveness of AVR techniques is to dynamically check how many of the candidate patches they produce are \textit{plausible}, i.e., they pass the relevant test cases after installing the patch in the project.
When a reliable test suite is unavailable, \textit{static metrics} are often employed, such as Exact Match or CodeBLEU~\cite{codebleu}.
These consist of checking whether candidate patches are ``similar'' to a given ground-truth patch, which is assumed to be the only correct way to repair that vulnerability.
Ground-truth patches are often derived from historical fixes committed to the project repository by human developers. 
AI-driven AVR techniques mostly adopt this type of assessment, as compiling each candidate patch and running test cases multiple times is ineffective for computing a loss function during training.

AVR techniques are intended to assist developers in repairing vulnerabilities by enabling them to \textbf{select} (i.e., choose) and refine plausible patches generated by one or more AVR tools on their own.
However, AVR techniques achieving strong performance on static and dynamic benchmarks do not guarantee that the patches align with developers’ expectations; namely, their perceived \textit{trustworthiness} and \textit{usefulness} of generated patches.
In a typical usage scenario, the AVR tool is launched on some vulnerable code, suggests one or more ``candidate'' patches (or ``plausible'' ones, if test cases were run), and the \textit{human reviewer} must inspect them and assess their suitability against some criteria (mainly correctness, but also readability and maintainability).
An ideal AVR tool should ensure that the patches it returns also align with developers' expectations, increasing the likelihood of acceptance and thereby confirming the tool's usefulness.
Unfortunately, there is currently a knowledge gap about what humans focus on when reviewing AVR-generated patches, especially those produced by generative AI (GenAI), e.g., Large Language Models (LLMs).

Prior work by Papotti et al.~\cite{papotti2024acceptance} observed that human reviewers are more willing to accept patches from a specialized security tool (AVR) than from a traditional Automated Program Repair (APR) tool, but did not investigate how patch characteristics influence human reviewers' selection. 
In general, human aspects are mostly neglected in current research in the field; a survey by Winter et al.~\cite{winter2022let} of 260 APR studies up to 2018 found that human studies on APR are rare, accounting for less than 7\% of the studies.
Among these, only four studies designed experiments in which participants rated the usefulness of a specific tool or its patches at a single point in the experiment, rather than as the main focus of the study.

This paper aims to understand \textbf{how humans perceive AI-generated security patches}.
In a typical scenario, an AVR tool is used to fix a security vulnerability, and developers obtain a set of patches from which they must select the one to apply to the source code.
Typically, this corresponds to the one the developers \textit{prefer}, i.e., the one they consider more useful and trustworthy.
Thus, the paper investigates the factors that influence which security patch (among a few) people choose to fix a vulnerability.

The first aspect we analyze is whether human developers tend to select patches they \textit{judge} to be better, e.g., because they understand their intent better or believe they fix the vulnerability correctly and without regression.
In particular, we check how much these judgments affect the patch they ultimately select.

\rqbox{1}{How do human \textbf{judgments} about individual patches affect the choice of which patch to use to fix a security vulnerability?}

The second aspect we analyze concerns the explanations that humans explicitly provide to motivate why the patch they chose was selected and preferred over others.

\rqbox{2}{What \textbf{reasons} drive humans' patch selection when fixing a security vulnerability?}

We hypothesize that humans may hold negative prejudices toward GenAI-generated patches, preferring human-authored patches when given the opportunity.
To better characterize the perceived trustworthiness of GenAI-generated patches, we examine how humans react when learning about the \textit{``attributed provenance''} of the patches, i.e., whether they were made by a human or by GenAI.
Specifically, we check whether they would reconsider the patch they originally selected.

\rqbox{3}{How do humans \textbf{react} after learning about the \textit{attributed provenance} of patches that fix a security vulnerability?}

Finally, we analyze humans' explicit comments once they learn about the provenance of the patches and how this relates to their reactions.

\rqbox{4}{What \textbf{opinions} do humans have about the \textbf{attributed provenance} of patches that fix a security vulnerabilities?}


To answer such questions, we conducted a \textbf{controlled experiment} with human participants who were tasked to (i) analyze a real-world vulnerability, (ii) inspect and evaluate two plausible patches generated by two different LLMs (\gpt and \claude), (iii) select the patch believed to be the most adequate to fix the vulnerability, and (iv) motivate their choice.
This activity was repeated for four different vulnerabilities.

After each selection, the participants were informed about the \textit{provenance} of the patches, stating that one originated from GenAI and the other from a human (which is a deliberately false fact), and they were allowed to reconsider their selection and justify it.
This aimed to understand their attitude toward GenAI, i.e., whether they were biased toward or against it.

The experiment involved voluntary participants from software security-related Master's courses at three universities: \textit{Hamburg University of Technology}, \textit{University of Trento}, and \textit{Vrije Universiteit Amsterdam}, all in Germany, Italy, and the Netherlands, respectively.
In total, \textbf{\acaAllParticipants Master's students} took part in the experiment.
The experiment was also replicated with \textbf{\indParticipants professionals} from three different industrial companies to analyze differences from the academic participants and, therefore, to provide a complementary perspective on this study.
The companies operate in the consultancy field and were \textit{FrontEndART}, \textit{Cefriel}, and \textit{Oppida}, from Hungary, Italy, and France, respectively.
We answer the research questions by first analyzing the two groups jointly and then analyzing each group independently, and we report noteworthy differences.

In summary, this paper makes the following contributions:
\begin{itemize}
    \item A \textbf{controlled experiment} with 125 participants (students and professionals) to identify the main drivers influencing the selection of patches for fixing security vulnerabilities.
    \item Evidence of \textbf{provenance-based bias}: participants show a negative bias (even when explicitly admitted) toward GenAI, leading them to change the initially selected patch if it is attributed to an LLM and to choose the human-made option regardless of technical merit.
    \item A set of \textbf{recommendations for AVR tool developers} to improve patch acceptance, prioritizing clear explanations, adherence to coding practices, and seamless integration into development workflows.
\end{itemize}

Data and scripts of this study are released for independent inspection and analysis (see Section \hyperref[sec:data:availability]{\textit{``Data Availability''}}).

\section{Experiment Design}
\label{sec:design}

\begin{figure}[t]
    \centering
    \includegraphics[width=1\linewidth]{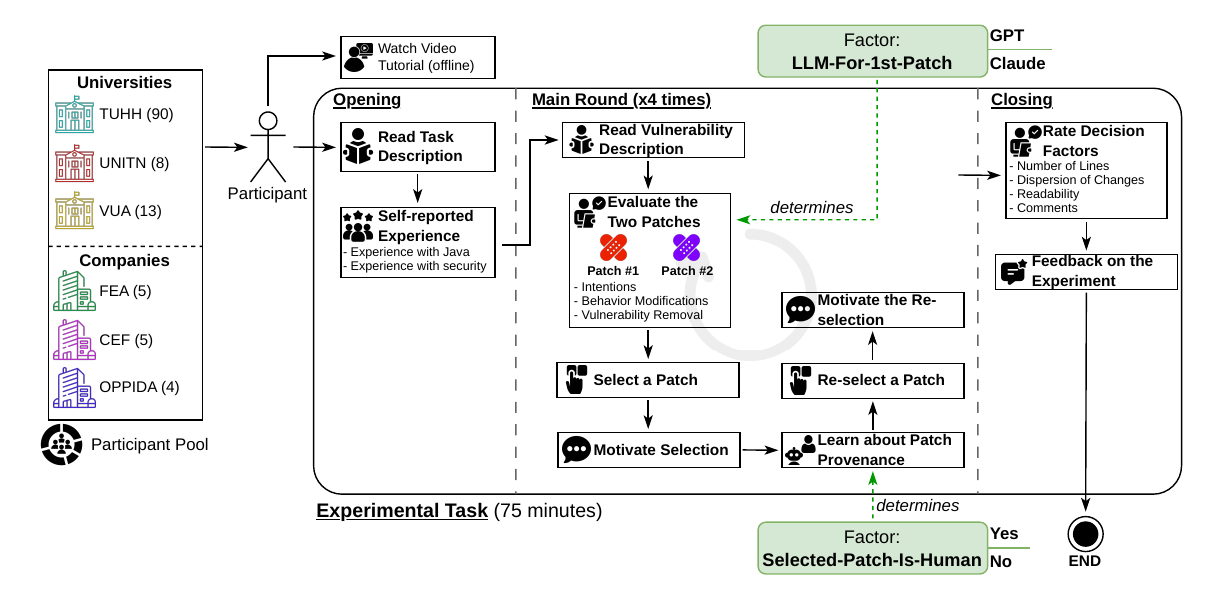}
    \caption{Overview of the experimental design.}
    \label{fig:experiment}
\end{figure}

Figure~\ref{fig:experiment} depicts the overall experimental procedure.
In the following sections, we explain the content of each part.

\subsection{Task Description}
\label{subsec:task}

The entire task consisted of filling in a \textit{questionnaire} within at most \textit{75 minutes}.
All participant responses were collected solely via the online questionnaire.
The questionnaire was administered via \textsc{LimeSurvey}, accessible via a web browser with an Internet connection.
The structure and content of the questionnaire are detailed in Table \ref{tab:questions}.

The experimental task is divided into three \textit{phases}:
\begin{enumerate}
    \item \textbf{Opening}: Participants were informed about the overall structure of the task and could provide some background information about themselves.
    \item \textbf{Main round} (x4 iterations): Participants were asked to analyze two plausible patches for a security vulnerability and select the most appropriate one in their opinion. They were also asked to reconsider their selection after the patches' attributed provenance was disclosed.
    \item \textbf{Closing}: Participants were asked questions concerning all four rounds and the whole experimental experience.
\end{enumerate}

We provide a detailed description of each section below.

\subsubsection{Opening}

The participants were welcomed and divided into the appropriate experimental groups, four in total.
The assigned group did not affect the overall structure of the experimental task, but rather some details, such as the order of the patches and the attributed provenance.
More details are provided in Section \ref{subsec:groups}.

At the start, participants were informed of the overall task structure, the rules to follow, and the maximum duration (75 minutes).
Before starting, we let the participants watch a short \textbf{video tutorial} (of 5 minutes) showing how to complete the questionnaire using an example vulnerability (that was not ultimately part of the real task).
This walkthrough acted as a training session.

\begin{table*}[t]
\centering
\caption{Questionnaire structure with the type of each answer.}
\label{tab:questions}
\resizebox{1\textwidth}{!}{%
    \begin{tabular}{|l|p{10cm}|l|}
    \hline
    \rowcolor{black}
    \textcolor{white}{\textbf{ID}} &
    \textcolor{white}{\textbf{Question}} & \textcolor{white}{\textbf{Answer Type}} \\
    \hline
    \rowcolor{gray20}
    \multicolumn{3}{|l|}{\textsc{\textbf{Opening}}} \\ \hline
    \ques{O1} & Please indicate your gender (optional): & Checkbox (opt.)$^*$ \\ 
    \ques{O2} & What is your experience with \textsc{Java} programming? & Multiple Choice$^{**}$ \\ 
    \ques{O3} & What is your experience with security vulnerabilities? & Multiple Choice$^{**}$ \\ 
    \hline
    \rowcolor{gray20}
    \multicolumn{3}{|l|}{\textsc{\textbf{Main Round} (Repeated four times, $i=1,2,3,4$)}} \\ \hline
    \rowcolor{blue!10}
    \multicolumn{3}{|l|}{Indicate how much you agree with the following statements concerning the \textbf{first patch}:} \\ \hline
    \rowcolor{blue!10}\ques[i]{R1} & I understand the intention of this patch. & 5-point Likert$^\dagger$ \\ 
    \rowcolor{blue!10}\ques[i]{R2} & I understand how this patch modifies the overall behavior of the code. & 5-point Likert$^\dagger$ \\ 
    \rowcolor{blue!10}\ques[i]{R3} & I understand how this patch removes the security vulnerability. & 5-point Likert$^\dagger$ \\ 
    \hline
    \rowcolor{blue!10}
    \multicolumn{3}{|l|}{Indicate how much you believe the following statements concerning the \textbf{first patch} are true:} \\ \hline
    \rowcolor{blue!10}\ques[i]{R4} & This patch successfully preserves the existing functionality of the code (irrespective of how it changes its security). & 5-point Likert$^\ddagger$ \\ 
    \rowcolor{blue!10}\ques[i]{R5} & This patch successfully removes the security vulnerability in the code (irrespective of it changes its existing functionality). & 5-point Likert$^\ddagger$ \\ 
    \rowcolor{blue!10}\ques[i]{R6} & This patch can be applied directly (i.e., no further changes are needed). & 5-point Likert$^\ddagger$ \\ 
    \hline
    \rowcolor{green!10}
    \multicolumn{3}{|l|}{Indicate how much you agree with the following statements concerning the \textbf{second patch}:} \\ \hline
    \rowcolor{green!10}\ques[i]{R7} & I understand the intention of this patch. & 5-point Likert$^\dagger$ \\ 
    \rowcolor{green!10}\ques[i]{R8} & I understand how this patch modifies the overall behavior of the code. & 5-point Likert$^\dagger$ \\ 
    \rowcolor{green!10}\ques[i]{R9} & I understand how this patch removes the security vulnerability. & 5-point Likert$^\dagger$ \\ 
    \hline
    \rowcolor{green!10}
    \multicolumn{3}{|l|}{Indicate how much you believe the following statements concerning the \textbf{second patch} are true:} \\ \hline
    \rowcolor{green!10}\ques[i]{R10} & This patch successfully preserves the existing functionality of the code (irrespective of how it changes its security). & 5-point Likert$^\ddagger$ \\ 
    \rowcolor{green!10}\ques[i]{R11} & This patch successfully removes the security vulnerability in the code (irrespective of whether it changes its existing functionality). & 5-point Likert$^\ddagger$ \\ 
    \rowcolor{green!10}\ques[i]{R12} & This patch can be applied directly (i.e., no further changes are needed). & 5-point Likert$^\ddagger$ \\ 
    \hline
    \ques[i]{R13} & In your view, which is the most appropriate patch for fixing the security vulnerability previously described? & \{First/left, Second/right\} \\ 
    \ques[i]{R14} & Please explain your choice (i.e., why you chose one over the other). & Free Text \\ 
    \hline
    \ques[i]{R15} & Please select again the most appropriate patch for fixing the security vulnerability previously described. & \{First/left, Second/right\} \\ 
    \ques[i]{R16} & Please explain the rationale behind your final decision. & Free Text \\ 
    \hline
    \rowcolor{gray20}
    \multicolumn{3}{|l|}{\textsc{\textbf{Closing}}} \\ \hline
    \multicolumn{3}{|p{15cm}|}{Indicate how much you agree with the following statements concerning all the patches you have seen throughout the entire experiment:} \\ \hline
    \ques{C1} & The number of lines added and deleted by the patches influenced my decisions. & 5-point Likert$^\dagger$ \\ 
    \ques{C2} & The degree of dispersion of the changes made by the patches influenced my decisions. & 5-point Likert$^\dagger$ \\ 
    \ques{C3} & The overall readability of the patches influenced my decisions. & 5-point Likert$^\dagger$ \\ 
    \ques{C4} & The comments added by the patches influenced my decisions. & 5-point Likert$^\dagger$ \\ 
    \hline
    %
    \multicolumn{3}{|l|}{Indicate how much you agree with the following statements concerning the whole experience:} \\ \hline
    \ques{C5} & I had a clear understanding of the whole experiment. & 5-point Likert$^\dagger$ \\ 
    \ques{C6} & I had enough time to complete the whole experiment. & 5-point Likert$^\dagger$ \\ 
    \ques{C7} & The information provided allowed me to complete the whole experiment confidently. & 5-point Likert$^\dagger$ \\ 
    \ques{C8} & Feel free to report any additional comments or remarks about this experiment. & Free Text (opt.) \\ 
    \hline
    \multicolumn{3}{l}{opt. $=$ Optional} \\
    \multicolumn{3}{p{15cm}}{$*$ None, one or more of: \{Woman, Man, Non-binary, Prefer not to say, Prefer to self-describe\}, with the latter a free text field.} \\
    \multicolumn{3}{p{15cm}}{$**$ One of: \{No experience, Attended a tutorial, Attended a course (e.g., university), Company internship, Professional practice\}.} \\
    \multicolumn{3}{l}{$^\dagger$ One of: \{Strongly Disagree, Disagree, Neither Agree nor Disagree, Agree, Strongly Agree\}.} \\
    \multicolumn{3}{l}{$\ddagger$ One of: \{Definitely no, Probably no, Not sure, Probably yes, Definitely yes\}.} \\
    \end{tabular}
}
\end{table*}

After this, the first page of the questionnaire recapped the previously provided information and asked for consent.
The participants were prompted to provide background information about themselves, i.e., their gender and prior experience with \textsc{Java} programming and security vulnerabilities (\ques{O1}--\ques{O3} in Table \ref{tab:questions}).

\subsubsection{Main Rounds}
\label{subsubsec:main_rounds}

After the opening, the participants entered a loop of \textit{four main rounds}, each presenting a different vulnerability to analyze, and two plausible patches generated by two distinct LLMs, i.e., \gpt and \claude (more details in Section~\ref{subsec:preparation}).
A vulnerability was described with a simplified issue description that explains the general type of weakness and provides a brief overview of the negative consequences of the specific vulnerability.
The source code of the affected methods is also given.
We deliberately did not point to the exact location of the vulnerability (i.e., the vulnerable statements) as this would bias participants' judgment of the patches; for instance, a patch that did not modify or delete all the vulnerable statements could have been improperly disregarded.

The participants were asked to review the two plausible patches, presented in \textit{unified diff} format, i.e., showing the lines added in green (marked with \texttt{+}) and the lines deleted in red (marked with \texttt{--}).
At this stage, the participants were still unaware of the provenance of the two patches and were only told that both were \textit{plausible}, i.e., are compilable and pass the relevant tests (more details in Section~\ref{subsec:preparation}).
The two patches are shown side by side as high-resolution screenshots captured in \textit{Visual Studio Code} (light mode), with their syntax highlighted.
However, the exact \textbf{order of display}, i.e., which appeared on the left and which on the right, was determined by the assigned experimental group (more details in Section~\ref{subsec:groups}).

Next, the participants were prompted to rate some aspects of each patch on a 5-point Likert scale by answering \textbf{six} questions per patch.
The first three concerned how much the participant \textit{understood} (i) the intention of the patch, (ii) how it modified the overall behavior of the code, and (iii) how it removed the security vulnerability.
The latter three questions concern how much the participant \textit{agreed} that (iv) the patch preserved the existing functionality, (v) it removed the vulnerabilities, and (vi) it could be applied without further changes in the project.
The exact text of these questions is reported in Table~\ref{tab:questions} (\ques[i]{R1}--\ques[i]{R6} for the first patch and \ques[i]{R7}--\ques[i]{R12} for the second patch).

Afterward, the participants were asked to \textbf{select} the patch they considered most suitable for fixing the vulnerability described (\ques[i]{R13} in Table~\ref{tab:questions}).
Even if both patches were considered adequate, participants could express only one preference and motivate their choice through a free-text field (\ques[i]{R14} in Table~\ref{tab:questions}).

Then, the participants were directed to a new page of the questionnaire and informed about the \textit{attributed provenance} of the two patches in the round.
Namely, they were told that the \textbf{selected patch} was made either by a human developer or by an LLM, depending on the participant’s experimental group (as detailed in Section~\ref{subsec:groups}).
Conversely, the non-selected patch was disclosed as originating from the other source.
In reality, all patches involved in this experiment were made by LLMs (either \gpt or \claude); this intentionally misleading information was intended to measure participants' prejudice against LLMs.
Indeed, after this reveal, the participants were given the option to reconsider their previous choice and \textbf{re-select the patch} (\ques[i]{R15} in Table~\ref{tab:questions}), and to explain their decision through a free-text field (\ques[i]{R16} in Table~\ref{tab:questions}).

\subsubsection{Closing}
\label{subsubsec:closing}

After completing all four rounds, participants answered questions about all eight patches seen across the rounds.
They were asked to rate (on a 5-point Likert scale) the extent to which some \textit{characteristics} of the patches may have influenced their decisions, i.e., (i) the \textbf{number of lines} added and deleted, (ii) the degree of \textbf{dispersion} of the changes, (iii) the overall \textbf{readability}~\cite{scalabrino:jsep18:readability}, and (iv) the \textbf{comments added} in the patches.
These rates were collected through questions \ques{C1}--\ques{C4} in Table~\ref{tab:questions}.

The final part of the questionnaire collected feedback on the overall experimental experience.
This included whether participants understood the experiment in its entirety, had sufficient time to complete the task, and felt that the information provided was adequate (\ques{C5}--\ques{C7} in Table~\ref{tab:questions}).
Participants were also given the opportunity to leave free-text comments or remarks (\ques{C8} in Table~\ref{tab:questions}), which we ultimately did not process or analyze in the present study.

After all responses were submitted, the participants were informed of the true provenance of the patches: none were written by human developers; all were generated by LLMs.

\subsection{Experimental Groups}
\label{subsec:groups}

This experiment manipulated \textbf{two independent variables} (a.k.a. \textit{factors}).
The first one, called \textbf{\llmForFirstPatch}, determines which patch to display in the first position (i.e., the left-hand side) based on the name of the LLM that generated it.
We hypothesized that the patch's position may influence participants' selection (e.g., a systematic preference for the first one).
Besides, it would mitigate bias among participants who systematically select the first (or second) patch in each round, thereby influencing our analyses.
This variable has two levels, i.e., $\{\abbrevgpt, \abbrevclaude\}$ (nominal scale), corresponding to \gpt and \claude, respectively.
A given level also determines the LLM for the second patch; e.g., \llmForFirstPatch$=\abbrevgpt$ implies that the first patch is generated by \gpt and the second by \claude.

The second independent variable, \textbf{\selectedIsHuman}, tells whether the provenance of the selected patch is attributed to a human developer (regardless of whether it was generated by \gpt or \claude).
Consequently, the other patch is attributed to an LLM.
This variable has two possible levels, i.e., $\{Y, N\}$ (nominal scale), denoting `Yes' and `No', respectively.
When \selectedIsHuman$=Y$, the selected patch is attributed to a human and the not-selected one to an LLM.
Conversely, when \selectedIsHuman$=N$, the selected patch is attributed to an LLM and the not-selected one to a human.

\begin{table}[t]
    \centering
    \caption{Crossover-balanced design to make four balanced experimental groups.}
    \label{tab:design}
    \begin{tabular}{cc|cccc}
        \toprule
        & & \multicolumn{4}{c}{\textbf{Round}} \\
        & & R1 & R2 & R3 & R4 \\
        \hline
        & Group-1 & $\langle\abbrevgpt,Y\rangle$ & $\langle\abbrevclaude,N\rangle$ & $\langle\abbrevclaude,N\rangle$ & $\langle\abbrevgpt,Y\rangle$ \\
        & Group-2 & $\langle\abbrevclaude,Y\rangle$ & $\langle\abbrevgpt,N\rangle$ & $\langle\abbrevgpt,N\rangle$ & $\langle\abbrevclaude,Y\rangle$ \\
        & Group-3 & $\langle\abbrevclaude,N\rangle$ & $\langle\abbrevgpt,Y\rangle$ & $\langle\abbrevgpt,Y\rangle$ & $\langle\abbrevclaude,N\rangle$ \\
        \multirow{-4}{*}{\textbf{\rotatebox{90}{Group}}}
        & Group-4 & $\langle\abbrevgpt,N\rangle$ & $\langle\abbrevclaude,Y\rangle$ & $\langle\abbrevclaude,Y\rangle$ & $\langle\abbrevgpt,N\rangle$ \\ 
        \bottomrule
        \multicolumn{6}{p{7cm}}{Each cell contains a pair $\langle x, y\rangle$, where $x$ = \llmForFirstPatch, while $y$ = \selectedIsHuman.}
    \end{tabular}
\end{table}

Since the experimental task comprises four rounds (one per vulnerability), exhaustive coverage of all treatments would require 256 unique combinations ($2^4 \times 2^4$), which would need a large number of participants.
We adopted a \textit{crossover-balanced design} as described by Massacci et al.~\cite{massacci2024addressing} to maximize the number of data points and ensure that the frequency of each independent variable assessment is balanced across the four rounds, without the need to cover all potential treatments.
More specifically, this is a repeated-measures design in which each participant performs the task on multiple distinct scenarios (here, four vulnerabilities) under different experimental configurations. With two levels for each independent variable and four rounds, each participant is exposed to each level of both variables exactly twice; across groups, each vulnerability is assessed once under each of the four treatment combinations. Moreover, the sequences are counterbalanced so that, for each variable, the number of times that one level precedes the other is equal to the number of times it appears in the reverse order, thereby mitigating potential order and learning effects.

This resulted in the selection of just \textbf{four treatments} ($\langle\abbrevgpt$, $Y\rangle$, $\langle\abbrevgpt$, $N\rangle$ $\langle\abbrevclaude$, $Y\rangle$ $\langle\abbrevclaude$, $N\rangle$), resulting in \textbf{four experimental groups} (Group $1$--$4$), which are summarized in Table~\ref{tab:design}.
This design ensures that all participants experience all levels of each independent variable at least once, regardless of their group assignment.
It also prevents any group (and its participants) from receiving the same treatment more than twice.

\subsection{Patch Preparation}
\label{subsec:preparation}

We selected two state-of-the-art (at the time of the experiment) commercial LLMs, \gpt and \claude, as both demonstrated strong repair performance and other security-related tasks~\cite{belozerov2026secure}.
The experimental task involved four real-world vulnerabilities affecting open-source \textsc{Java} projects, presented in a fixed order to all participants, regardless of assigned group.

Given a vulnerability, we generated the patches by prompting the two LLMs using the strategy defined by Antal et al.~\cite{antal2025identifying}.
Namely, for each vulnerability, we constructed a prompt containing (i) the source code of the affected method, (ii) some contextual information such as the class header, its fields, and the signatures of methods referenced within the vulnerable method, and (iii) the vulnerability description.
We removed the original code comments from all presented code elements (class declaration, method declaration, field declaration) to prevent the model from relying on them, as they might be misleading (e.g., an outdated description~\cite{sabetto2025impact}).
Furthermore, not every vulnerable method had such comments, so removing them ensured all code snippets were evaluated under the same conditions.

The two LLMs were called with a temperature of $0.01$ to minimize output variability and ensure a more deterministic and reproducible repair process, consistent with prior automated vulnerability repair studies.
For each vulnerability, candidate patches were generated in a \textit{zero-shot} setting (i.e., without model fine-tuning, iterations, or few-shot examples), reflecting a typical scenario in which developers would use LLMs.

We selected the four vulnerabilities from the Vul4J+~\cite{vul4jplus2024dataset} dataset, containing, among others, 106 vulnerabilities that can be reproduced with executable \textbf{vulnerability-witnessing tests}, i.e., \textsc{JUnit} test cases that fail on the vulnerable version of the code but pass on the patched version, acting as a ``dynamic'' oracle of the LLM-generated patches.
We first selected vulnerabilities that (i) were localized in a single file to facilitate inspection by participants and (ii) covered different weakness types to increase representativeness.
From all vulnerabilities that matched these criteria, we randomly sampled one at a time and called both LLMs to generate candidate patches.
We selected the pair of patches (one per LLM) such that (i) they could be placed in the project and make the whole project build successfully, (ii) they passed all the vulnerability-witnessing tests related to the vulnerabilities as in Vul4J+ (i.e., were \textit{plausible}), and (iii) exhibited some syntactic and semantic differences (i.e., they were not equivalent).
If any patches did not meet these conditions, they were discarded and regenerated by re-prompting the LLM.
Furthermore, the plausible patches were manually reviewed to confirm that they could fix the vulnerability without altering the original behavior.
If, after \textit{five} attempts, the LLMs were not able to produce the desired patches, a new vulnerability was selected. 

\begin{table}[t]
\small
\centering
\caption{Objective characteristics of the eight patches shown to participants.}
\label{tab:patch_characteristics}
\resizebox{\linewidth}{!}{%
\begin{tabular}{llrrrrrr}
\toprule
\textbf{Vulnerability} & \textbf{Made By} & \textbf{Lines Add.} & \textbf{Lines Del.} & \textbf{Total Churn} & \textbf{Comm. Add.} & \textbf{Hunks} \\
\hline
& \gptShort & 24 & 1 & 25 & 8 & 8 \\
\multirow{-2}{*}{VUL4J-5} & \claudeShort & 17 & 1 & 18 & 4 & 6 \\
\hline
& \gptShort & 45 & 4 & 49 & 9 & 2 \\
\multirow{-2}{*}{VUL4J-12} & \claudeShort & 58 & 8 & 66 & 11 & 2 \\
\hline
& \gptShort & 20 & 5 & 25 & 6 & 6 \\
\multirow{-2}{*}{VUL4J-13} & \claudeShort & 20 & 6 & 26 & 4 & 6 \\
\hline
& \gptShort & 38 & 9 & 47 & 10 & 4 \\
\multirow{-2}{*}{VUL4J-64} & \claudeShort & 75 & 8 & 83 & 9 & 3 \\
\bottomrule
\multicolumn{7}{l}{Comm. $=$ Comments, Add. $=$ Added, Del. $=$ Deleted}
\end{tabular}
}
\end{table}

The vulnerabilities ultimately selected were the following:

\begin{enumerate}
    \item \textbf{VUL4J-5} (from APACHE-COMMONS-001, affecting Apache Commons Compress) is a path traversal vulnerability that occurs during unzipping crafted ZIP archives (a.k.a. Zip Slip, CWE-23). This case occurs when the target directory has a sibling directory, and the target directory's name is a prefix of the sibling's name.
    \item \textbf{VUL4J-12} (from CVE-2018-17202, affecting Apache Commons Imaging) is an infinite loop vulnerability (CWE-835) caused by the loop variable in the affected method that was not updated properly, leading to a non-terminating program with crafted JPEG files.
    \item \textbf{VUL4J-13} (from CVE-2018-17201, affecting Apache Commons Imaging) is an infinite loop vulnerability (CWE-835) caused when a certain \texttt{InputStream} is empty, leading the program to spin on a non-empty stream indefinitely with crafted RGBE files.
    \item \textbf{VUL4J-64} (from CVE-2018-20157, affecting OpenRefine) is an XML external entity injection (XXE) vulnerability (CWE-611) caused by crafted XML documents. This specific case occurs when the input XML document contains a Document Type Definition (DTD) with URIs to sensitive files, causing the XML parser to embed their content into the XML document.
\end{enumerate}

Table~\ref{tab:patch_characteristics} summarizes key facts of the eight patches shown to the participants.
The number of modified lines varied across vulnerabilities, ranging from concise fixes to more extensive modifications that introduced additional validation logic or helper methods.
Consequently, they added more lines than they deleted, as typically expected from vulnerability fixes.
The number of comments added by the patches also differed, with some patches containing little documentation and others including more detailed explanations.
In contrast, the changes were generally narrowly dispersed, as all patches modified one code region and therefore consisted of a single diff hunk~\cite{alali:icpc2008:hunk}.
These characteristics correspond to the factors that participants were later asked to reflect upon when reporting which aspects of the patches influenced their decisions.

\subsection{Participant Selection and Response Screening}

The experiment was carried out in two contexts: one comprising Master's students enrolled in university courses (\textit{academics}), and one comprising industrial companies (\textit{professionals}).

\subsubsection{Academics}

The participants were students enrolled in Master’s programs in computer science (or closely related fields) at three universities.
The students attended specialized courses in software security, where they had previously attended classes on common vulnerabilities and how to detect and avoid them.
Thus, they all had sufficient background to take part in the experiment.
Specifically, we recruited students from the courses of:
\begin{itemize}
    \item \textit{Software Security}, \textit{Software Security Engineering}, and \textit{Cybersecurity Data Science} at the \textbf{Hamburg University of Technology} (TUHH), Germany.
    \item \textit{Security Experiments: Attacks and Defenses} at the \textbf{University of Trento} (UNITN), Italy.
    \item \textit{Security Testing} at the \textbf{Vrije Universiteit Amsterdam} (VUA), in The Netherlands.
\end{itemize}

The students were selected on a \textbf{voluntary} basis (i.e., participation was not mandatory) after a public notice was posted through the courses' communication channels and verbally during previous classes.
To encourage participation, we awarded students \textit{bonus points} for completing the experimental task and demonstrating serious involvement (see Section~\ref{subsec:data_analysis} for more details).
Overall, we recruited 90 students from TUHH, 8 from UNITN, and 13 from VUA; this gap was due to differences in the number of students enrolled in those courses.

The participant provided their responses anonymously.
Namely, we asked them to provide a \textit{participant ID}, a pseudonymized identifier computed solely by the participant and nearly impossible to trace back to the person~\cite{leo:er2008:identifier}.
This had the sole goal of assigning bonus points correctly; indeed, participants were asked to provide their ID after the course final grade was issued so it could be increased by the bonus points.
The participants provided their ID on the first page of the questionnaire.
Each participant generated their ID according to the following rules:
\begin{enumerate}
    \item First letter of the mother's first name (e.g., ``Linda'' $\xrightarrow{}$ \texttt{L}).
    \item Day of birth (2 digits) (e.g., ``5 April 1996'' $\xrightarrow{}$ \texttt{05}). 
    \item Last letter of the birth place (e.g., ``Berlin'' $\xrightarrow{}$ \texttt{N}).
    \item First letter of the father's first name (e.g., ``Robert'' $\xrightarrow{}$ \texttt{R}).
    \item Last digit of the university matriculation number (e.g., ``1234567'' $\xrightarrow{}$ \texttt{7}).
\end{enumerate}

Due to differing course schedules, we ran the experiment in three separate sessions: one at the end of May 2025 (for TUHH students), one in September 2025 (for UNITN students), and one in January 2026 (for VUA students).
On the day of the experimental session, participants were asked to provide formal \textit{consent} to use their responses for the intended research purpose (i.e., to analyze them).
This choice did not affect the assignment of bonus points---i.e., a participant could participate in the experiment only for the sake of the bonus points.
All this information was reported in a \textbf{consent form} distributed to prospective participants before they joined the experiment.
We release a copy of the form in our replication package~\cite{appendix}.

\subsubsection{Professionals}

We compared the results of academics with professionals from three industrial companies, involving a total of \indParticipants participants (5, 5, and 4, respectively):
\begin{itemize}
    \item \textit{FrontEndART} is a Hungarian software company specializing in code analysis, software quality assurance, and research-driven software engineering tools, with strong ties to academic research in software security. The five participating software developers have at least 5 years of industrial experience; 3 hold a PhD, and the remaining 2 have more than 10 years of experience in industrial Java development.
    \item \textit{Cefriel} is an Italian not-for-profit digital innovation center founded by the Politecnico di Milano, specializing in research, innovation, and education in digital technologies. Its core areas of expertise include cybersecurity and custom software engineering, leveraging cutting-edge technologies to address clients' unique challenges across industries.
    The participants in the experiment were all software developers with 10+ years of experience.
    \item \textit{Oppida} is a company specialized in cybersecurity audits. It is an accredited evaluation body for cybersecurity (ITSEF), which grants it international recognition for its expertise. The participants in the experiment are all vetted ITSEF evaluators, comprising 3 experts (5+ years of experience) and 1 junior, with expertise in pentesting, cryptography analysis, and code review.
\end{itemize}

Participants were recruited voluntarily through an internal selection process conducted within each company.
The experimental task, design, and materials were identical to those described in Sections~\ref{subsec:task}--\ref{subsec:preparation}, except that the experiment was conducted remotely rather than in person, as it would be difficult to fit all participants' schedules.

Despite the large (anticipated) difference in sample size relative to the academic population, we preserved the division into the four experimental groups and distributed participants randomly and uniformly across them.

\subsubsection{Valid Responses}

We received 136 complete responses in total.
We first excluded 17 responses from participants who did not authorize the processing of their data for research purposes.

Within the questionnaire, we included two simple \textit{attention check} questions for each round, one per patch, for a total of eight questions.
These were intended to identify participants who did not engage seriously with the task.
These questions asked participants to identify straightforward facts about each patch, such as the total number of methods in the resulting code or the name of a newly introduced method.
The exact set of questions is reported in the paper's online appendix~\cite{appendix}.
We excluded the responses from participants who failed the majority of these questions, i.e., at least five.
This led us to discard 26 responses.
Ultimately, we retained 93 valid responses: 59 from TUHH, 7 from UNITN, 13 from VUA, and \indParticipants from industrial companies.
Thus, the academic sample accounts for 85\% of the pool.

\subsubsection{Demographics}
\label{subsub:demographics}

\begin{figure}[t]
    \centering
    \begin{subfigure}[b]{0.49\linewidth}
        \centering
        \includegraphics[width=\linewidth]{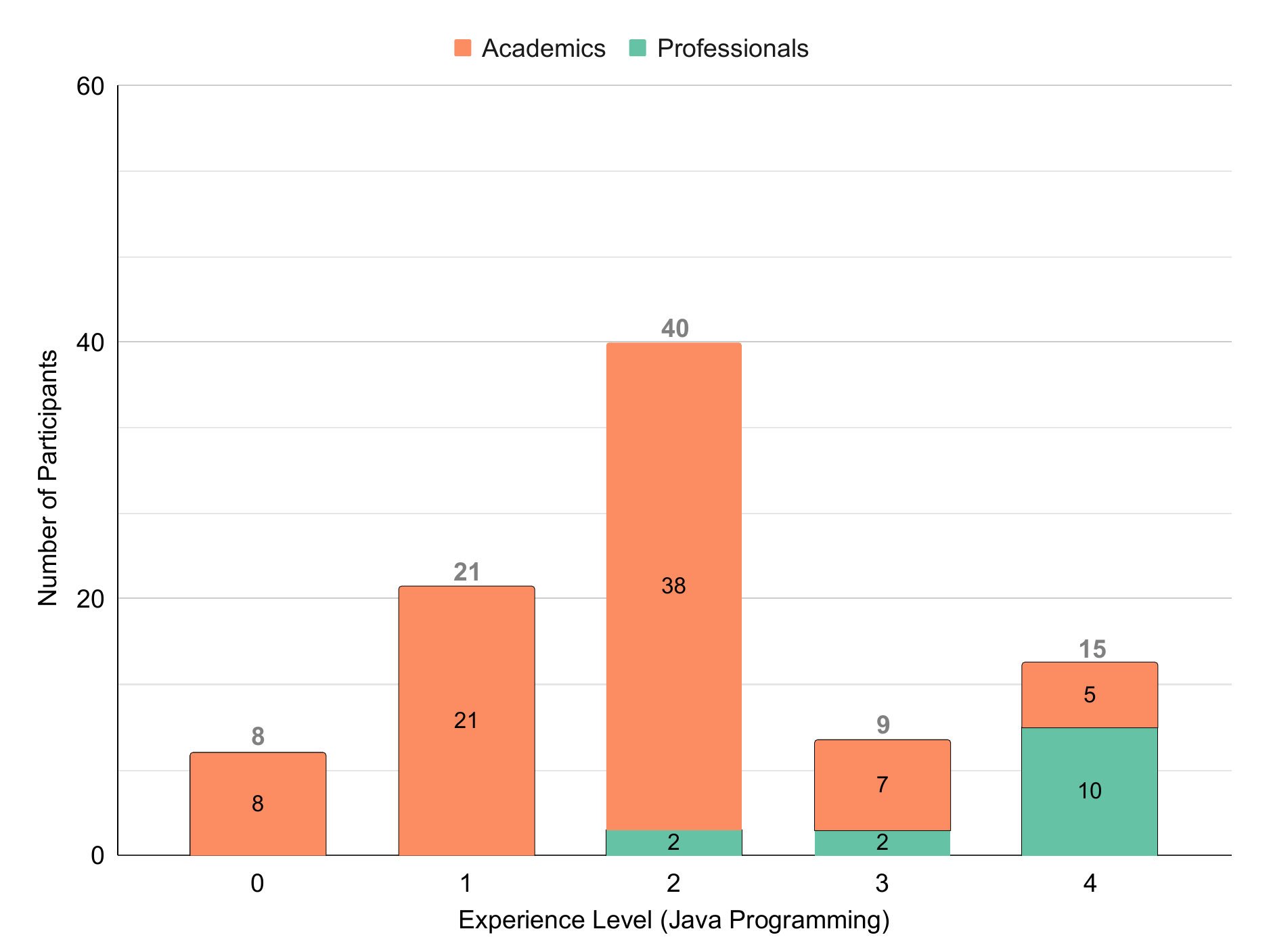}
        \caption{Experience with Java programming.}
        \label{fig:exp_java}
    \end{subfigure}
    \hfill
    \begin{subfigure}[b]{0.49\linewidth}
        \centering
        \includegraphics[width=\linewidth]{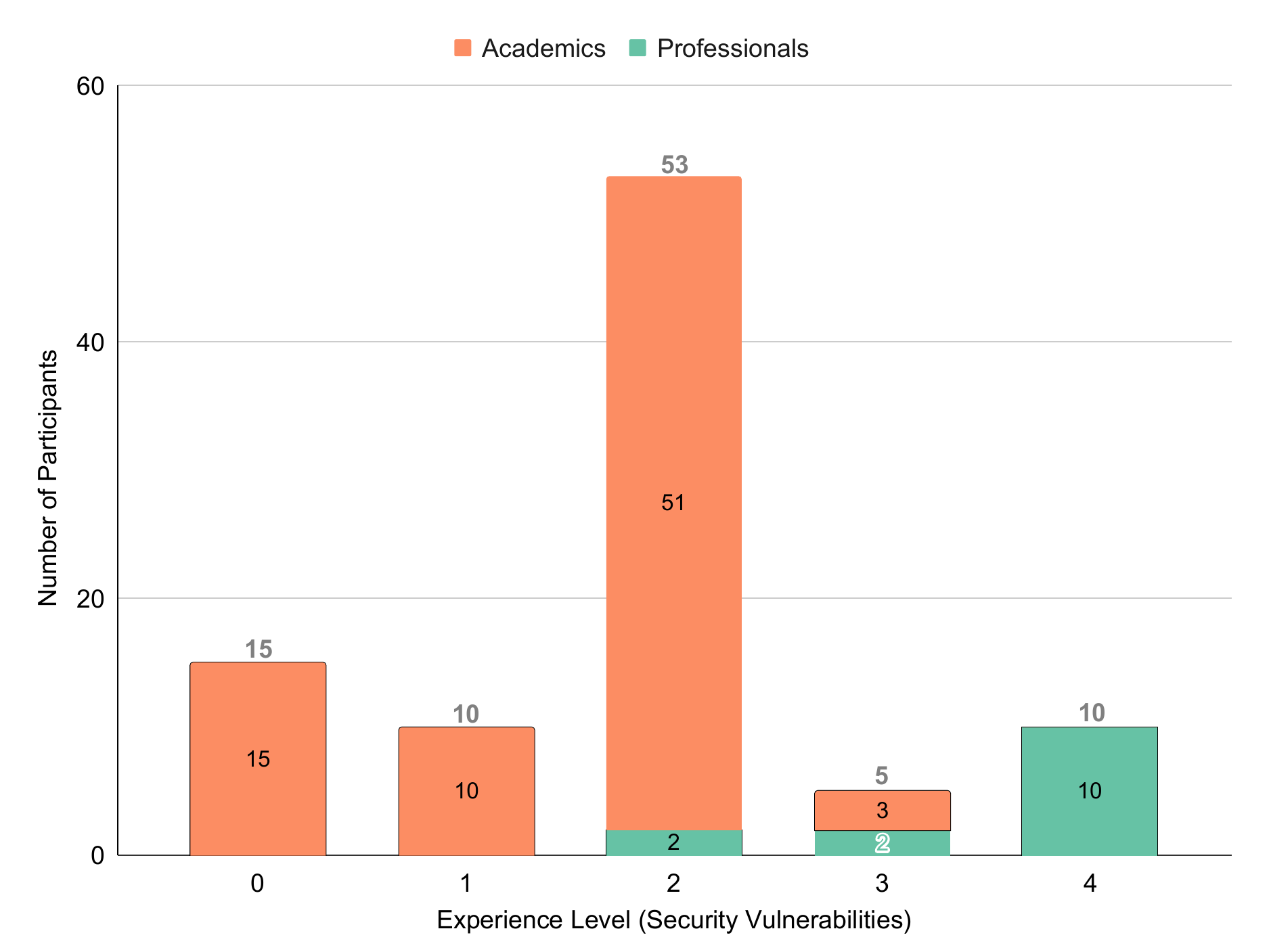}
        \caption{Experience with security vulnerabilities.}
        \label{fig:exp_security}
    \end{subfigure}
    \caption{Number of participants based on their self-disclosed experience level (N=\allValidParticipants).}
    \label{fig:participant_experience}
\end{figure}

Figure~\ref{fig:participant_experience} shows the distribution of participants across the self-disclosed experience levels in Java programming and security vulnerabilities, ranging from 0 (no experience) to 4 (professional practice).
Regarding experience in Java, 8 (8.60\%), 21 (22.58\%), 40 (43.01\%), 9 (9.68\%), and 15 (16.13\%) participants reported levels 0 to 4, respectively.
The majority of participants reported a middle level of experience (2), corresponding to ``Attended a course'', followed by levels 1, 4, 3, and 0.
The overall distribution is heavily weighted toward academic participants, who comprise 85\% of the pool.
However, we observe that professionals reported markedly higher levels (as expected).
Specifically, 10 of the 14 professionals indicated maximum experience, while the remaining four reported levels 2 and 3.

Regarding the experience with security vulnerabilities, 15 (16.13\%), 10 (10.75\%), 53 (56.99\%), 5 (5.38\%), and 10 (10.75\%) participants reported levels 0 to 4, respectively.
The majority of participants reported a middle level of experience (2), corresponding to ``Attended a course'', followed by levels 0, 1, and 4 (equally frequent as 1), with level 3 the least common.
Here too, the global distribution is influenced by the larger number of academic participants (85\% of the pool).
The distribution within professionals is exactly like the one for Java programming: 10 reported maximum experience, and the other four selected levels 2 and 3.

\subsection{Data Analysis Protocol}
\label{subsec:data_analysis}

\subsubsection{Judgments Affecting the Selection (\rqq{1})}
\label{subsub:rq1_design}

In Section~\ref{subsubsec:main_rounds}, we described the six questions that capture participants' judgments on various aspects of each patch (questions \ques[i]{R1}--\ques[i]{R12} in Table~\ref{tab:questions}).
Each judgment can be expressed through a 5-point Likert scale (ordinal).
We assigned identifiers to each of the following aspects: $Int$ (\ques[i]{R1} and \ques[i]{R7}), $Mod$ (\ques[i]{R2} and \ques[i]{R8}), $Fix$ (\ques[i]{R3} and \ques[i]{R9}), $Prs$ (\ques[i]{R4} and \ques[i]{R10}), $Rem$ (\ques[i]{R5} and \ques[i]{R11}), and $Dir$ (\ques[i]{R6} and \ques[i]{R12}).

Then, we analyzed the extent to which these judgments influenced the patch that participants selected across the rounds \textit{before the attributed provenance is disclosed}.
To this end, we fit a \textbf{logistic regression} model to identify which of these aspects affected patch selection.
Specifically, we fit three models: one encompassing all participants, one for academic participants, and one for professional participants.
The models' predictors encoded the difference between the judgments assigned to the two patches.
Namely, the judgments to the $i$-th patch were encoded as $\bm{Int_i}$, $\bm{Mod_i}$, $\bm{Fix_i}$, $\bm{Prs_i}$, $\bm{Rem_i}$, $\bm{Dir_i}$ (all on an ordinal scale).
Then, we computed the differences that were used as models' predictors: $\bm{\Delta Int}$=$Int_2$-$Int_1$, $\bm{\Delta Mod}$=$Mod_2$-$Mod_1$, $\bm{\Delta Fix}$=$Fix_2$-$Fix_1$, $\bm{\Delta Prs}$=$Prs_2$-$Prs_1$, $\bm{\Delta Rem}$=$Rem_2$-$Rem_1$, and $\bm{\Delta Dir}$=$Dir_2$-$Dir_1$.
The models' dependent variable was $\bm{Sel}$, which represented the patch a participant selected in one round (through \ques[i]{R13} in Table~\ref{tab:questions}).
Specifically, we encoded it as a binary variable with value $0$ if the first patch was selected and $1$ if the second one was selected.
More technically, the model estimates how the difference in the judgments of the two patches (second patch minus the first) affects the probability of selecting the second patch.
Given the four rounds, each participant's response produced four distinct observations, for a total of \allObservations, \acaObservations, and \indObservations data points for the three models, respectively.

We assessed the significance of the estimated predictor coefficients.
We further analyzed predictors with statistically significant coefficients ($p< 0.05$) by examining their \textit{odds ratios} (ORs) to quantify the strength of the association between the predictor and the dependent variable.
That is, a one-unit increase in the predictor (i.e., an extra point on the rating) increases the odds of the outcome by the OR, provided that all other variables remain unchanged.

\subsubsection{Reasons for Selection (\rqq{2})}
\label{subsubsec:rq2_design}

We systematically analyzed participants' answers to question \ques[i]{R14} (Table~\ref{tab:questions}) to distill the reasons (i.e., motivations) they explicitly stated (or hinted at) for selecting patches across the four main rounds ($i=1,2,3,4$).
These answers concern the selection participants make before the patches' attributed provenance is disclosed in each round.
To this end, we performed a \textbf{thematic analysis}~\cite{guest2011applied,Cruzes:esem2011:thematic:se} of the \allObservations free-text answers (from \allValidParticipants participants) provided across all four rounds.

The process began with \textbf{inductive open coding}, assigning one or more \textit{codes} to each answer that describe \textit{reasons for selection}.
This activity was carried out by a co-author of this paper who labeled a sample of 100 answers ($\approx$27\% of the total).
The inspector did this initial coding without a predefined list of codes to avoid biasing judgment.
At this stage, the inspector kept granularity low, i.e., identifying codes based on the text and avoiding overinterpretation.
After the first pass, the inspector produced 137 low-level codes.

All codes were then reviewed and aggregated to eliminate overlaps and inconsistencies.
Then, each refined code was given a \textit{definition}, and codes with logical relationships were grouped into coherent \textit{themes}.
This resulted in a first version of the \textbf{codebook}.
Afterward, the inspector made a second pass on the sample of 100 answers to update the assigned codes according to the established codebook.

During the inspection of this sample, the inspector noted that some answers justified \textit{the rejection of the non-selected patch} rather than justifying the selection of the chosen patch.
In these instances, the reasoning relied on a process of \textit{elimination} rather than an evaluation of the selected patch's quality.
This occurred in 11 out of 100 cases.
For example, in answer \#325, the explanation focused on the shortcomings of the second patch (\answerVerbatim{The second patch still seems to have a security vulnerability. It tries to filter paths by looking for... but that can probably be bypassed}).
To maintain coding consistency and avoid conflating two distinct groups of codes (one addressing the chosen patch and the other the rejected alternatives), the inspector framed all codes \textit{in terms of the selected patches}.
In answer \#325, the assigned code was ultimately \codetheme{Fix the Vulnerability Adequately} because the second patch was rejected for failing to adequately address the vulnerability, implicitly communicating that the selected patch was chosen because it successfully resolved the issue.

To speed up coding the remaining 272 answers, the inspector used \deepseekLong to generate preliminary code proposals for the remaining answers, given the previously established codebook.
The model was also given a sample of 100 previously coded answers to help it understand how to code like the inspector.
The model was also allowed to propose new codes for the codebook if necessary; none were ultimately proposed.
The exact prompt is provided in Appendix~\ref{appendix:prompt_rq2}.
Then, the inspector reviewed the model results and corrected the wrongly assigned or missing codes.
In doing so, the inspector further refined some code definitions and produced a second version of the codebook (reported in Section~\ref{subsec:rq2_results}).

After the first inspector finished coding, we asked another co-author to \textit{double-check} the codes assigned by the main inspector.
We randomly sampled a new set of 100 answers out of \allObservations and asked the co-author to verify that (1) the assigned codes were correct and (2) new codes that were missing could be added, i.e., it was possible to propose new codes or refine existing ones.
Then, we computed the \textit{validation rate}, i.e., the proportion of answers for which the second inspector confirmed the validity of the assigned codes.
The resulting rate was 92\% ($\pm$ 5.32, 95\% CI), indicating 8 initial disagreements.
The two inspectors discussed these cases until they reached agreement, which led to adding missing codes to 3 cases, though this did not require updating the code list or its definitions.

Once all answers were coded, we analyzed recurring themes and codes by examining their absolute counts and global frequencies (relative to all answers).
We also examined cases in which the answer explicitly indicated that the participant was unsure which of the two patches to select.

Then, we compared the distribution of themes and codes across the two participant samples (i.e., academics vs. professionals) to check for tendencies to mention a concept more or less often.
In particular, we performed a \textit{chi-square test of independence} for each code and theme to assess whether the differences between academics and professionals were statistically significant.
This resulted in several hypothesis tests, requiring adjustment of the resulting p-values using the Benjamini-Hochberg procedure to control the false discovery rate~\cite{bh:correction}.

In Section~\ref{subsubsec:closing}, we described the four questions that capture participants' judgments on further aspects of the patches that may have influenced their decisions across all rounds (questions \ques{C1}--\ques{C4} in Table~\ref{tab:questions}).
Specifically, (i) the \textit{number of added/deleted lines}, (ii) the \textit{degree of dispersion of changes}, (iii) the \textit{overall readability}, and (iv) the \textit{comments added} in the patches.
We checked whether the ratings (on a 5-point Likert scale) for these aspects aligned with the identified codes and themes, e.g., whether there were inconsistencies.
Specifically, we examined the mean scores and the number of participants who reported a high score (4 or 5).
Each response provided a single observation, as participants answered these questions only once, for a total of \allValidParticipants (\acaValidParticipants academics and \indParticipants professionals).

Furthermore, we divided the participants by the self-reported experience levels in Java programming and security vulnerabilities (from questions 
\ques{O2} and \ques{O3} in Table~\ref{tab:questions}, respectively) and compared the resulting subgroups.
Specifically, we examined whether code distributions differed across experience levels.
To this end, we performed an omnibus chi-square test of independence for each code, resulting in multiple statistical tests; the resulting p-values were adjusted using the Benjamini-Hochberg procedure to control the false discovery rate~\cite{bh:correction}.
For codes rejecting the tests' null hypothesis (adjusted $p<0.05$), we conducted a post-hoc analysis by examining the \textit{adjusted standardized residuals} of the contingency table, which quantify the difference between the observed and expected frequencies for each cell of the table, expressed in approximate units of standard deviation~\cite{Haberman:residuals}. 
Thus, a large positive residual indicates that the code occurred more frequently than expected under the assumption of independence, whereas a large negative residual indicates that it occurred less frequently than expected.
We computed the residuals $|z_j|$ for each experience level ($j=0,1,2,3,4$) and inspected cases where $|z_j| > 2.576$. 
This threshold corresponds to a significance level of $\alpha = 0.01$ for a two-tailed test and was obtained by applying a Bonferroni correction to account for the multiple comparisons performed in the post-hoc analysis.

\subsubsection{Reaction on Attributed Provenance (\rqq{3})}

We analyzed participants' decisions to \textit{keep} or \textit{switch} the initially selected patch after disclosing the attributed provenance (\ques[i]{R15} in Table~\ref{tab:questions}) to assess whether they preferred human-authored patches.
Depending on the assigned experimental group, either the selected patch or the non-selected patch was attributed to the human source (see Section~\ref{subsec:groups} for details).

We first counted how many times participants kept the selected patch or switched to the other one (i.e., they changed their initial decision) across the four rounds.
In particular, we counted how many times these two events occurred when the selected patch was attributed to an LLM (corresponding to \selectedIsHuman$=N$) and when it was attributed to a human (corresponding to \selectedIsHuman$=Y$).
We made this count for the academic and professional subgroups as well, resulting in three contingency tables that relate the factor \selectedIsHuman to the decision to keep or switch (binary).

Then, we performed a chi-squared test of independence for each contingency table to assess whether \textit{attribution to a human or an LLM significantly influences the decision to keep or switch}, to test the hypothesis that humans prefer human-made patches over LLM-generated patches.
Given the three tests, we applied a Bonferroni correction to control the family-wise error rate, adjusting the p-values for multiple comparisons by multiplying each raw p-value by the number of tests (i.e., 3).
We considered adjusted p-values below $0.05$ statistically significant, which confirms that the two variables are related.
In such cases, we also computed the \textbf{odds ratio} to quantify the strength of association.
An odds ratio greater than 1 indicates that participants are more likely to switch their decision when the selected patch was attributed to an LLM rather than a human.

\subsubsection{Opinions on Attributed Provenance (\rqq{4})}

We systematically analyzed participants’ answers to question \ques[i]{R16} (Table~\ref{tab:questions}) to distill participants' opinions about the disclosed provenance attributed to the patches across the four main rounds ($i=1,2,3,4$).
We followed a similar procedure to that in \rqq{2}, i.e., a co-author of this paper manually conducted a thematic analysis~\cite{guest2011applied,Cruzes:esem2011:thematic:se} of the \allObservations free-text answers (from \allValidParticipants participants) provided across all four rounds.
They started with a sample of 100 answers ($\approx$27\% of the total); however, unlike \rqq{2}, the inspector first checked whether participants \textit{explicitly acknowledged the provenance}.
Indeed, we hypothesized that participants might not have commented on provenance and instead focused on the patches themselves.
Specifically, the inspector checked whether the answers (i) expressed a general opinion on humans and LLMs or (ii) explained how this information about the provenance had an effect on their decision process.
This means that an answer that used the terms ``human'' or ``LLM'' just to refer to the two patches would not confirm acknowledgment.

For the answers where there was no acknowledgment, the inspector also checked the respective first answers given before the provenance disclosure (to question \ques[i]{R14}) and classified the second answer into three classes:
\begin{itemize}
    \item \textit{Restate Original Motivations}, if the second answer merely restates the motivations already given in the first one, even in a summarized way.
    \item \textit{Provide Additional Motivations}, if the second answer provides some more motivations than those given in the first one.
    \item \textit{Unclear/Unspecified Motivations}, in all other cases, e.g., when the second answer gives no clear motivations.
\end{itemize}

For answers that acknowledged provenance, the inspector performed inductive open coding by assigning one or more \textit{codes} to each answer that describe the \textit{opinions of participants in light of the provenance}, e.g., motivations explaining why they kept their selection or switched, or other general comments about humans and LLMs
This initial coding was done without a predefined list of codes to avoid biasing the judgment.
At this stage, the inspector kept granularity low, i.e., identifying codes based on the text and avoiding overinterpretation.

After the first pass, the inspector produced 46 low-level codes, which were further reviewed to eliminate overlaps and inconsistencies.
Each refined code was given a \textit{definition}, and codes with logical relationships were grouped into coherent \textit{themes}.
This resulted in a first version of the \textbf{codebook}.
Afterward, the inspector made a second pass on the sample of 100 answers to refine the assigned codes according to the established codebook.

To speed up coding the remaining 272 answers, the inspector used \deepseekLong to generate preliminary codes.
The model was asked to do the very same task as the inspector: determine whether provenance was acknowledged and, if not, assign the right class (out of three); if yes, assign codes according to the established codebook.
As in \rqq{2}, the model was given the sample of previously labeled answers for guidance and was allowed to propose new codes if necessary (none was ultimately proposed).
The exact prompt is provided in Appendix~\ref{appendix:prompt_rq4}.
The inspector reviewed the model results and corrected the wrongly assigned or missing codes.
This led the inspector to refine some code definitions and produce a second version of the codebook (reported in Section~\ref{subsec:rq4_results}).

As with \rqq{2}, we also involved another co-author to \textit{double-check} the codes assigned by the main inspector.
We randomly sampled a new set of 100 answers out of \allObservations and asked the co-author to verify whether (1) the provenance is (or is not) acknowledged in the answer, (2) the assigned class or codes were correct, and (3) new codes that were missing could be added, i.e., it was possible to propose new codes or refine existing ones.
We computed the \textit{validation rate} accordingly: 96\% ($\pm$ 3.84, 95\% CI), indicating 4 initial disagreements.
The two inspectors discussed these cases until they reached agreement, which led to changing the class for 2 answers that did not acknowledge provenance; thus, this did not require updating the code list or its definitions.

Once all answers were coded, we counted how many explicitly acknowledged provenance.
First, we counted how answers with no acknowledgment were distributed across the three classes.
We also compared the distribution of such answers across the two participant samples (i.e., academics vs. professionals) to check for specific tendencies.
Then, for answers with an acknowledgment, we analyzed the distribution of themes and codes as in \rqq{2}, including comparisons between academic and professional subgroups.

Furthermore, we assessed whether answers reflected participants' ultimate decision to \textit{keep} their originally selected patch or \textit{switch} to the other one (already analyzed in the context of \rqq{3}).
Specifically, we investigated whether participants who did \textit{not} acknowledge the attributed provenance tended to \textit{keep} their originally selected patch.
Indeed, we hypothesized that participants would be less likely to switch patches if they did not comment explicitly on the provenance.
Practically, we categorized the answers into two groups corresponding to each decision type (``Switch'' vs. ``Keep'') and compared them using chi-square tests of independence to test whether the provenance acknowledgment differed significantly between the two groups.
Since we ran one test per participant sample, we adjusted the resulting p-values with the Benjamini-Hochberg procedure to control the false discovery rate~\cite{bh:correction}.
In cases where the test was rejected (adjusted $p< 0.05$), we quantified the strength of the association using \textit{odds ratios} (ORs).
An odds ratio greater than 1 indicates that participants are more likely to switch their decision when the attributed provenance is explicitly commented on. 

\section{Result Analysis}
\label{sec:results}

\subsection{Judgments Affecting the Selection (\rqq{1})}
\label{subsec:rq1_results}

Table~\ref{tab:logreg} summarizes three fitted logistic regression models that use participants' judgments on the six patch aspects to predict which of the two patches was selected ($Sel$, as defined in Section~\ref{subsub:rq1_design}).
All three models were statistically significant ($p<0.05$) and achieved a pseudo-$R^2$ of at least 0.30, indicating a \textit{substantial} improvement in fit compared to intercept-only models.
This suggests that the differences in judging the two patches (i.e., the $\Delta$ variables) can be used to predict which patch will be selected.

\begin{table}[t]
\caption{Results of the fitted logistic regression models predicting patch selection ($Sel$) in a round. The table reports estimated regression coefficients, standard errors (SE), $p$-values, and odds ratios (ORs) for each predictor, defined as the difference in judgments between the two patches in a round (second minus first). Statistically significant predictors ($p < 0.05$) are shown in \textbf{bold}.}
\label{tab:logreg}
\resizebox{\linewidth}{!}{%
\begin{tabular}{l|rrrr|rrrr|rrrr}
\hline
\multirow{2}{*}{\textbf{Predictor}} & \multicolumn{4}{c|}{\textbf{Academics (N=\acaObservations)}} & \multicolumn{4}{c|}{\textbf{Professionals (N=\indObservations)}} & \multicolumn{4}{c}{\textbf{All (N=\allObservations)}} \\
& \textbf{Coeff.} & \textbf{SE} & \textbf{$p$} & \textbf{OR} & \textbf{Coeff.} & \textbf{SE} & \textbf{$p$} & \textbf{OR} & \textbf{Coeff.} & \textbf{SE} & \textbf{$p$} & \textbf{OR} \\
\hline
(Intercept) & 0.035 & 0.148 & 0.811 & 1.036 & -0.201 & 0.373 & 0.590 & 0.818 & 0.019 & 0.135 & 0.889 & 1.019 \\
$\Delta Int$ & -0.003 & 0.308 & 0.993 & 0.997 & 12.166 & 297.909 & 0.967 & $>$1,000$^*$ & 0.148 & 0.296 & 0.617 & 1.160 \\
$\Delta Mod$ & 0.453 & 0.285 & 0.113 & 1.572 & 0.714 & 1.945 & 0.713 & 2.043 & 0.425 & 0.275 & 0.123 & 1.530 \\
$\Delta Fix$ & \textbf{1.042} & \textbf{0.305} & \textbf{0.001} & \textbf{2.836} & -1.510 & 1.771 & 0.394 & 0.221 & \textbf{0.921} & \textbf{0.292} & \textbf{0.002} & \textbf{2.512} \\
$\Delta Prs$ & \textbf{1.025} & \textbf{0.264} & \textbf{0.000} & \textbf{2.788} & 0.590 & 0.634 & 0.352 & 1.804 & \textbf{0.970} & \textbf{0.238} & \textbf{0.000} & \textbf{2.637} \\
$\Delta Rem$ & \textbf{0.655} & \textbf{0.203} & \textbf{0.001} & \textbf{1.925} & 1.316 & 0.996 & 0.186 & 3.728 & \textbf{0.719} & \textbf{0.199} & \textbf{0.000} & \textbf{2.053} \\
$\Delta Dir$ & \textbf{0.555} & \textbf{0.190} & \textbf{0.003} & \textbf{1.742} & 1.309 & 0.999 & 0.190 & 3.704 & \textbf{0.616} & \textbf{0.181} & \textbf{0.001} & \textbf{1.851} \\
\hline
& \multicolumn{4}{p{3cm}|}{McFadden $R^2 {=} 0.300$; $p {=} 6.87e-26$} & \multicolumn{4}{p{3cm}|}{McFadden $R^2 {=} 0.419$; $p {=} 1.47e-05$} & \multicolumn{4}{p{3cm}}{McFadden $R^2 {=} 0.303$; $p {=} 3.8e-31$} \\ 
\hline
\multicolumn{13}{l}{$^*$OR of $\Delta Int$ in the model for \textit{Professionals}: $192,214.259$}
\end{tabular}
}
\end{table}

Specifically, in the global model (which includes all \allObservations observations), four predictors were significant: $\Delta Fix$, $\Delta Prs$, $\Delta Rem$, $\Delta Dir$.
Thus, participants primarily selected patches based on (i) perceived preservation of existing functionality, (ii) understanding of the security fix, (iii) belief that the patch removes the vulnerability, and (iv) confidence in its applicability without further changes.
In particular, the positive coefficients indicate that patches receiving higher scores on these aspects were more likely to be selected.
The strongest effects were observed for $\Delta Prs$ and $\Delta Fix$ by analyzing the respective odds ratios (OR): a one-point increase in their difference (score given to the second patch minus the first one) increased the odds of selecting the second one by approximately 2.6 and 2.5, respectively.
The odds ratios for $\Delta Rem$ and $\Delta Dir$ were approximately 2.1 and 1.9, respectively, indicating a similar effect as the other predictors.

These findings largely persist when analyzing only the \acaObservations observations from academic participants, with minor differences in the exact odds-ratio values but similar overall effects.
However, the model fitted to \indObservations observations from professional participants revealed no significant predictors, providing insufficient evidence to draw conclusions for this subgroup.
We also observed that the OR for predictor $\Delta Int$ was unusually large, nearly 200,000.
This number does not indicate a real effect, but suggests sparse data.

\rqanswer{1}{
Participants primarily selected a patch based on their judgment of four aspects: (i) the extent to which it preserves existing functionality, (ii) how well they understood its security fix, (iii) the extent to which they believed it removes the vulnerability, and (iv) whether it can be applied without further changes.
Between two patches, the patch receiving the higher score on these aspects is more likely to be selected; a one-point increase corresponds to at least a twofold increase in the odds of selection.
}

\subsection{Reasons for Selection (\rqq{2})}
\label{subsec:rq2_results}

\subsubsection{The Recurring Themes and Codes}

Table~\ref{tab:rq2_codes} reports the codes and themes resulting from the thematic analysis of answers to question \ques[i]{R14} (Table~\ref{tab:questions}).
The table also shows how many times each code and theme appeared in participants' answers, along with their relative frequencies across all \allObservations answers (called ``global frequency'') and within answers belonging to their theme (called ``in-theme frequency'').

As anticipated, the most common theme across all participants was \codetheme{Security}: almost half of the answers (45\%) cited the patch's security correctness as a reason for selection, i.e., whether it fixed the vulnerability more adequately, validated input more thoroughly, or handled permissions more safely.
Specifically, the participants explained that the chosen patch appeared better suited to the vulnerability than the other patch (\codetheme{Fix the Vulnerability Adequately}), i.e., the fix implemented is likely more correct than the other patch's.
In some cases, participants went into more detail, commenting on how well the input validation (\codetheme{Validates input Adequately}) or the permission-handling mechanisms (\codetheme{Better Permission Handling}) were implemented.
These codes directly stem from the type of vulnerability the participants were given, as they mostly could be fixed with sanitization or by adjusting permissions.
In some cases, they provided more generic statements about the patch being more reliable, safe, or secure, but without explicitly commenting on the fix's correctness (\codetheme{Generally More Reliable/Safe/Secure}).
We observed that security correctness is a fundamental requirement: patches that did not address the vulnerability were not selected, even when they had other desirable properties, e.g., better comprehensibility.

The second most prevalent theme was \codetheme{Code Comprehension}: at least one out of three answers (36\%) indicated that the selection was also driven by the patch's ease of understanding.
Patches that were more readable (\codetheme{Generally More Understandable}), introduced more comments (\codetheme{More or Better Comments}), and/or had an overall better coding structure (\codetheme{Better Code Structure and Organization}) were particularly appreciated and seem to largely influence the chance of being selected.

The third most prevalent theme was \codetheme{Robustness and Functional Behavior}, concerning how well the patch handles the various input types (e.g., malformed ones) and does not break the existing functional behavior of the code being fixed.
In almost one out of three answers (31\%), participants expressed preference for patches that employed a better error handling mechanism, e.g., with more try-catch blocks (code \codetheme{Better Error Handling}) or handled more input-dependent logic by adding new conditional statements (code \codetheme{Handles More Cases}), as long as they did not disrupt existing behavior (code \codetheme{Preserve Functionality}).
Thus, patches are selected more often if they anticipate malformed input and error states, but without harming the existing logic.

The fourth most prevalent theme was \codetheme{Maintainability}, with its concepts mentioned in almost one in five answers (20\%).
Within this, the most important aspect participants appreciated was the patch's modularization of the source code (\codetheme{Modularizes Code}), i.e., it created separate methods to handle specific logic, such as dedicated methods for input sanitization or error handling.
This also partially influenced code comprehension, as modular code improves structure and organization (\codetheme{Better Code Structure and Organization}, part of theme \codetheme{Code Comprehension}); however, the participants did not always explicitly state that separate methods improve comprehension, which motivated our decision to include \codetheme{Modularizes Code} under the \codetheme{Maintainability} theme.
Similarly to comprehension, patches that are easier to maintain seem more likely to be selected.

Lastly, we observed that at least one answer out of ten (10\%) provided very generic or unclear explanations (theme \codetheme{Generic or Unspecified Reasons}).
Namely, the participants merely explained that a patch was selected because it was better than the other without providing examples or further details.
In a few cases, the patch was selected because it did something some participants appreciated, e.g., avoiding bit-level arithmetic, but without explaining why this choice negatively affects a quality aspect, such as comprehension or maintainability.
In 14 cases, the answer did not provide any valid reason, for example, by stating that the choice was made randomly, or by expressing uncertainty.

\clearpage
{\tiny
\begin{longtable}{p{1.6cm}p{3.3cm}|r>{\raggedleft\arraybackslash}p{0.8cm}|r>{\raggedleft\arraybackslash}p{0.8cm}|r>{\raggedleft\arraybackslash}p{0.8cm}}
\caption{Codebook of the free answers to question \ques[i]{R14} ($i=1,2,3,4$) motivating the selection of the chosen patch, with the number of answers (\#) containing each code (white rows) and theme (gray rows). The column ``Freq.'' reports the global frequency (outside parentheses) and in-theme frequency (in parentheses).}\label{tab:rq2_codes}\\
\hline\hline
\multirow{2}{=}{\textbf{Code/Theme Name}} &
\multirow{2}{=}{\textbf{Description (`The patch was selected...')}} & \multicolumn{2}{>{\centering\arraybackslash}p{1.45cm}|}{\textbf{Academics (N=\acaObservations)}} & \multicolumn{2}{>{\centering\arraybackslash}p{1.45cm}|}{\textbf{Professionals (N=\indObservations)}} & \multicolumn{2}{>{\centering\arraybackslash}p{1.45cm}}{\textbf{All (N=\allObservations)}} \\
& & \textbf{\#} & \textbf{Freq.} & \textbf{\#} & \textbf{Freq.} & \textbf{\#} & \textbf{Freq.} \\

\hline
\endhead

\multicolumn{8}{r}{\textit{Continued on next page}} \\
\endfoot

\endlastfoot

\rowcolor{gray20}
\textbf{Code Comprehension} & For how comprehensible the patch was (in absolute terms or relative to the other patch), mainly describing the changes that made the code more comprehensible, or stating that it was generally perceived as more comprehensible. & \textbf{109} & \textbf{34.49\%}\par(\textbf{-}) & \textbf{24} & \textbf{42.86\%}\par(\textbf{-}) & \textbf{133} & \textbf{35.75\%}\par(\textbf{-}) \\ \cline{2-8}
Generally More Understandable & Because it is generally deemed as more understandable (e.g., more readable, easier to follow, more transparent, clearer, more precise, more straightforward, better written, simpler) than the other patch. & 54 & 17.09\% (49.54\%) & 10 & 17.86\% (41.67\%) & 64 & 17.20\% (48.12\%) \\ \cline{2-8}
More or Better Comments & Because it has more comments and/or higher quality comments (e.g., more detailed, concise, structured) that makes it clearer than the other patch. & 24 & 7.59\% (22.02\%) & 5 & 8.93\% (20.83\%) & 29 & 7.80\% (21.80\%) \\ \cline{2-8}
Better Code Structure and Organization & Because it has an overall better structure and organization of the source code than the other patch. & 27 & 8.54\% (24.77\%) & 1 & 1.79\% (4.17\%) & 28 & 7.53\% (21.05\%) \\ \cline{2-8}
Good Coding Practices & Because it follows better good coding practices (e.g., proper variable naming) than the other patch and/or the other patch does not (e.g., it uses hardcoded strings). & 10 & 3.16\% (9.17\%) & 8 & 14.29\% (33.33\%) & 18 & 4.84\% (13.53\%) \\ \cline{2-8}
Clearer Errors & Because its exceptions/errors are easier to understand (e.g., they have more informative messages) than the other patch. & 10 & 3.16\% (9.17\%) & 2 & 3.57\% (8.33\%) & 12 & 3.23\% (9.02\%) \\ \cline{2-8}
Less Code & Because it has fewer lines of code than the other patch. & 4 & 1.27\% (3.67\%) & 1 & 1.79\% (4.17\%) & 5 & 1.34\% (3.76\%) \\ \hline

\rowcolor{gray20}
\textbf{Maintainability} & For how maintainable the patch was (in absolute terms or relative to the other patch), mainly describing the changes that made the code more maintainable, or stating that it was generally perceived as more maintainable. & \textbf{59} & \textbf{18.67\%}\par(\textbf{-}) & \textbf{14} & \textbf{25.00\%}\par(\textbf{-}) & \textbf{73} & \textbf{19.62\%}\par(\textbf{-}) \\ \cline{2-8}
Modularizes Code & Because it introduces dedicated functions/methods (e.g., helpers), separates concerns, and/or is more modular than the other patch. & 28 & 8.86\% (47.46\%) & 6 & 10.71\% (42.86\%) & 34 & 9.14\% (46.58\%) \\ \cline{2-8}
Avoid Redundancy & Because it does not introduce redundant code (e.g., unnecessary functions/methods) and/or does not perform unnecessary changes. & 12 & 3.80\% (20.34\%) & 5 & 8.93\% (35.71\%) & 17 & 4.57\% (23.29\%) \\ \cline{2-8}
Appropriate Variable Modifiers & Because it sets more appropriate variable modifiers (e.g., final) than the other patch or avoids doing so. & 12 & 3.80\% (20.34\%) & 1 & 1.79\% (7.14\%) & 13 & 3.49\% (17.81\%) \\ \cline{2-8}
Generally More Maintainable & Because it is generally deemed more maintainable than the other patch (also as a result of some unspecified refactoring or reduced complexity). & 11 & 3.48\% (18.64\%) & 2 & 3.57\% (14.29\%) & 13 & 3.49\% (17.81\%) \\ \hline

\rowcolor{gray20}
\textbf{Security} & For how secure the patch was (in absolute terms or relative to the other patch), mainly describing the changes that made the code more secure, or stating that it was generally perceived as more secure. & \textbf{154} & \textbf{48.73\%}\par(\textbf{-}) & \textbf{15} & \textbf{26.79\%}\par(\textbf{-}) & \textbf{169} & \textbf{45.43\%}\par(\textbf{-}) \\ \cline{2-8}
Fix the Vulnerability Adequately & Because it fixes the vulnerability adequately or anyway more adequately than the other patch. & 73 & 23.10\% (47.40\%) & 6 & 10.71\% (40.00\%) & 79 & 21.24\% (46.75\%) \\ \cline{2-8}
Validates Input Adequately & Because it performs input validation (i.e., checking conformance to a format or constraints) adequately or anyway more adequately than the other patch. & 45 & 14.24\% (29.22\%) & 4 & 7.14\% (26.67\%) & 49 & 13.17\% (28.99\%) \\ \cline{2-8}
Better Permission Handling & Because it sets the right permissions (e.g., specific or default ones). & 18 & 5.70\% (11.69\%) & 3 & 5.36\% (20.00\%) & 21 & 5.65\% (12.43\%) \\ \cline{2-8}
Perform Multiple Checks & Because it implements the defense-in-depth principle by performing multiple checks (i.e., validating the input more than once). & 12 & 3.80\% (7.79\%) & 5 & 8.93\% (33.33\%) & 17 & 4.57\% (10.06\%) \\ \cline{2-8}
Sanitizes Input Adequately & Because it performs input normalization and/or sanitization (i.e., transforming the input into a safe format). & 15 & 4.75\% (9.74\%) & 2 & 3.57\% (13.33\%) & 17 & 4.57\% (10.06\%) \\ \cline{2-8}
Generally More Reliable/Safe/Secure & Because it is generally deemed as more reliable, safe, and/or secure than the other patch. & 18 & 5.70\% (11.69\%) & 0 & 0.00\% (0.00\%) & 18 & 4.84\% (10.65\%) \\ \cline{2-8}
Broader Vulnerability Fix & Because it fixes a vulnerability beyond the main one and/or adds an extra security feature. & 5 & 1.58\% (3.25\%) & 0 & 0.00\% (0.00\%) & 5 & 1.34\% (2.96\%) \\ \hline

\rowcolor{gray20}
\textbf{Robustness and Functional Behavior} & For how robust the patch was (in absolute terms or relative to the other patch), mainly by describing the broader and better handling of cases and errors, whether it preserved the original behavior, or stating that it was generally perceived as more robust. & \textbf{101} & \textbf{31.96\%}\par(\textbf{-}) & \textbf{14} & \textbf{25.00\%}\par(\textbf{-}) & \textbf{115} & \textbf{30.91\%}\par(\textbf{-}) \\ \cline{2-8}
Better Error Handling & Because it has a better exception/error handling mechanism (e.g., raising and catching an exception explicitly) or anyway better than the other patch. & 51 & 16.14\% (50.50\%) & 7 & 12.50\% (50.00\%) & 58 & 15.59\% (50.43\%) \\ \cline{2-8}
Preserve Functionality & Because it preserves (or does not break) the functional behavior of the original code. & 19 & 6.01\% (18.81\%) & 6 & 10.71\% (42.86\%) & 25 & 6.72\% (21.74\%) \\ \cline{2-8}
Handles More Cases & Because it handles (i.e., covers, catches, raises) more cases (including edge and corner cases) and/or takes more precautions than the other patch. & 22 & 6.96\% (21.78\%) & 1 & 1.79\% (7.14\%) & 23 & 6.18\% (20.00\%) \\ \cline{2-8}
Better Resource Handling & Because it handles the resources better (e.g., avoids resource leaks by closing a stream on exit) than the other patch. & 7 & 2.22\% (6.93\%) & 1 & 1.79\% (7.14\%) & 8 & 2.15\% (6.96\%) \\ \cline{2-8}
Generally More Robust & Because it is generally deemed as more robust than the other patch. & 6 & 1.90\% (5.94\%) & 0 & 0.00\% (0.00\%) & 6 & 1.61\% (5.22\%) \\ \hline

\rowcolor{gray20}
\textbf{Generic or Unspecified Reasons} & For the participant's personal preference for what the patch did, general positive feelings toward the patch, and any unclear explanations. & \textbf{32} & \textbf{10.13\%}\par(\textbf{-}) & \textbf{7} & \textbf{12.50\%}\par(\textbf{-}) & \textbf{39} & \textbf{10.48\%}\par(\textbf{-}) \\ \cline{2-8}
Generally Better & Because it is generally deemed as more complete, functional, reasonable, and effective than the other patch. & 16 & 5.06\% (50.00\%) & 2 & 3.57\% (28.57\%) & 18 & 4.84\% (46.15\%) \\ \cline{2-8}
Perform a Preferred Action & Because it makes a certain appreciated action (e.g., an arithmetic operation) that the other patch does not or avoids doing unappreciated actions (e.g., bit-level operations). & 5 & 1.58\% (15.63\%) & 2 & 3.57\% (28.57\%) & 7 & 1.88\% (17.95\%) \\ \cline{2-8}
Unspecified Reason & Without clear reasons (e.g., the participant said 'I don't know', 'they are the same', or explained the difference between the two patches).
& 11 & 3.48\% (34.38\%) & 3 & 5.36\% (42.86\%) & 14 & 3.76\% (35.90\%) \\ \hline\hline
\end{longtable}
}

In a subsequent analysis, we examined cases where answers explicitly indicated a conflict between the patches, i.e., the participant was unsure and had to make a \textit{trade-off}.
This phenomenon was observed in seven cases in particular. 
Notably, in answer \#341 (\answerVerbatim{While the second patch is easier to understand it misses one crucial step and that is the sanitation of the entryname. The first patch checks if the filename is valid, and does not allow for any sketchy path names and that they fufill the requirements}) the participant stated that while patch B was easier to understand, patch A was preferred as it validates the \texttt{filename} adequately, thus having greater security than patch B.
Consequently, this answer was assigned the code \codetheme{Validates Input Adequately}.
Similarly, in answer \#263 (\answerVerbatim{Its quite hard for me to say since the first code uses the close method which i quite like but it does not have as strict conditions as the second code which is why i chose the second snippet}) the participant noted that patch A provided a more understandable and appreciated solution, but patch B was preferred as it performed stricter checks.
Accordingly, this answer was also assigned the code \codetheme{Validates Input Adequately}.
In answer \#188 (\answerVerbatim{Even though I like about the first patch candidate that the comments are more abundant and explain the behavior of the code better to someone not knowledgeable with the code (or Java, for that matter), I chose the second patch candidate. Though lacking in comment explanation, it seems to me that both patch candidates attempt to solve the issue the same way (by disabling external entity processing). [...] As someone not versed in the intricacies at play there, this seems more reliable and trustworthy to me. In fact, I am not sure whether the first patch candidate would even work/compute.}) the participant said that patch A explained the code better thanks to the amount of comments, but they preferred patch B as it looked more reliable and trustworthy (thus, the code assigned was \codetheme{Generally More Reliable/Safe/Secure}), without explaining too much why.
This in-depth analysis further supports the finding that participants prioritize correctness (from a security standpoint) as the primary factor driving selection, whereas readability is preferred but not decisive.


The distribution of codes and themes among academics aligns with the overall distribution, given their larger number compared with professional participants.
All the facts presented so far hold for the academic sample as well.
However, the professional participants follow a slightly different distribution of codes and themes when considered alone.
Notably, the theme \codetheme{Security} was the second most widely used, rather than the first, with \codetheme{Code Comprehension} leading.
Specifically, professionals appeared overall less interested than academics in how accurately the selected patch fixed the vulnerability (code \codetheme{Fix the Vulnerability Adequately}, -12\%), and, by extension, far less interested in explicitly commenting on security aspects in general (theme \codetheme{Security}, -22\%).
Conversely, they showed greater interest in adopting good coding practices (code \codetheme{Good Coding Practices}, +11\%) and in code comprehension more broadly (theme \codetheme{Code Comprehension}, +9\%), even if they had less interest in code structuring (code \codetheme{Better Code Structure and Organization}, -7\%).
To a lesser extent, they also paid more attention to the absence of redundant code (+5\%) and \codetheme{Maintainability} in general (+6\%).
The results from the 30 chi-square tests of independence (one per code and theme) confirmed the large differences in code \codetheme{Good Coding Practices} and theme \codetheme{Security}, as the corresponding adjusted p-values were below $0.05$.
Thus, our results suggest that, within our participant sample, professionals place greater emphasis on good coding practices than academics, while placing less emphasis on general security aspects.

One explanation for these results is that professional participants took the patch's security and correctness for granted, trusting more in the pre-assessment the patches had (indeed, all patches were plausible), leading them to focus on other quality aspects.
At the same time, this may also be influenced by their experience with Java and security vulnerabilities: 10 of the 14 participants reported the highest (level 4) in both.
Given the small sample size and the predominance of level-4 professionals, we cannot meaningfully analyze how codes and themes are distributed across all five experience levels for them alone.

\subsubsection{Relationship with Factors Influencing Decisions}

After completing all four rounds, participants were asked to rate four factors that may have influenced their overall decision process (\ques{C1}--\ques{C4} in Table~\ref{tab:questions}).
Specifically, (i) the \textit{number of added/deleted lines}, (ii) the \textit{degree of dispersion of changes}, (iii) the \textit{overall readability}, and (iv) the \textit{comments added} in the patches.
Table~\ref{tab:factors_influencing} summarizes their ratings.

The overall readability of the patches was reported as the most influential factor, followed by the comments the patches introduced.
Both aspects are related to patch understandability.
This confirms the evidence from the qualitative analysis of participants' answers, in which \codetheme{Code Comprehension} is the second-most frequent theme overall.
In contrast, the number of added and deleted lines, as well as the dispersion of changes, was reported to have less influence on the selection process.
The former is confirmed by the limited number of occurrences of the code \codetheme{Less Code} (only five times across all answers), implying that the size of the patch (in terms of lines changed) does not play an important role in the evaluation (at least in the eight patches of this experiment).
The latter, instead, is consistent with the number of hunks the two patches shown at each round had (see Table~\ref{tab:patch_characteristics}).
Indeed, the largest gap was 2 in the VUL4J-5 round, in which one patch modified 6 hunks while the other modified 8.
In the VUL4J-64 round, the gap was just one (4 vs. 3), while in the other two rounds the two patches had no differences in the number of hunks.
Thus, each patch pair showed similar variation in the dispersion of changes.

\begin{table}[t]
\caption{Self-reported scores of influence (from 1 to 5) of factors about patches on the selection made by the participants.}
\label{tab:factors_influencing}
\resizebox{\linewidth}{!}{%
\begin{tabular}{l|rl|rl|rl}
\hline
\multirow{2}{*}{\textbf{Factor}} & \multicolumn{2}{c|}{\textbf{Academics (N=79)}} & \multicolumn{2}{c|}{\textbf{Professionals (N=14)}} & \multicolumn{2}{c}{\textbf{All (N=93)}} \\
& \textbf{Mean Score} & \textbf{Score $\geq$4} & \textbf{Mean Score} & \textbf{Score $\geq$4} & \textbf{Mean Score} & \textbf{Score $\geq$4} \\
\hline
Number of added/deleted lines & 3.05 & 34 (43.04\%) & 3.07 & 6 (42.86\%) & 3.05 & 40 (43.01\%) \\
Degree of dispersion of changes & 3.38 & 42 (53.16\%) & 3.14 & 6 (42.86\%) & 3.34 & 48 (51.61\%) \\
Overall readability & 4.20 & 69 (87.34\%) & 4.07 & 11 (78.57\%) & 4.18 & 80 (86.02\%) \\
Comments added & 4.10 & 66 (83.54\%) & 3.71 & 10 (71.43\%) & 4.04 & 76 (81.72\%) \\
\hline
\end{tabular}
}
\end{table}

\subsubsection{Relationship with Experience Levels}

Furthermore, we related the codes to the \allValidParticipants participants' self-reported experience with Java programming and security vulnerabilities (see Section~\ref{subsub:demographics}), which were measured from 0 (no experience) to 4 (professional practice).
We present the analysis across all participants (academics and professionals combined) as academics substantially outnumber professionals, making the global distribution very similar to the academic-only one.
Additionally, 10 of the 14 professionals reported the maximum experience level in both dimensions, so analyzing them separately was less informative.
The plots concerning the individual samples can be found in the paper's online appendix~\cite{appendix}.

\begin{figure}[t]
    \centering
    \begin{subfigure}[b]{0.49\linewidth}
        \centering
        \includegraphics[width=\linewidth]{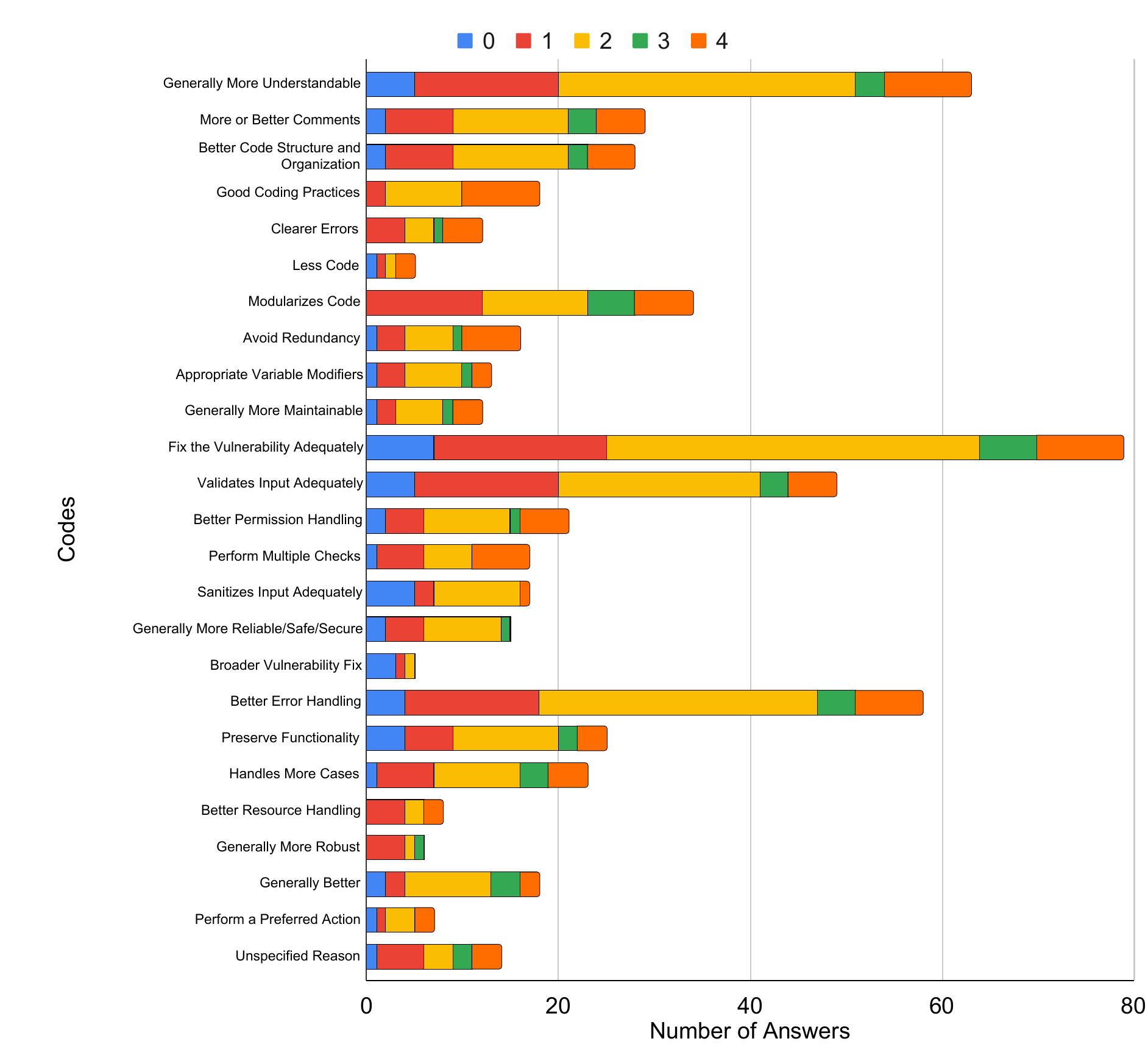}
        \caption{Experience with Java programming.}
        \label{fig:code_java}
    \end{subfigure}
    \hfill
    \begin{subfigure}[b]{0.49\linewidth}
        \centering
        \includegraphics[width=\linewidth]{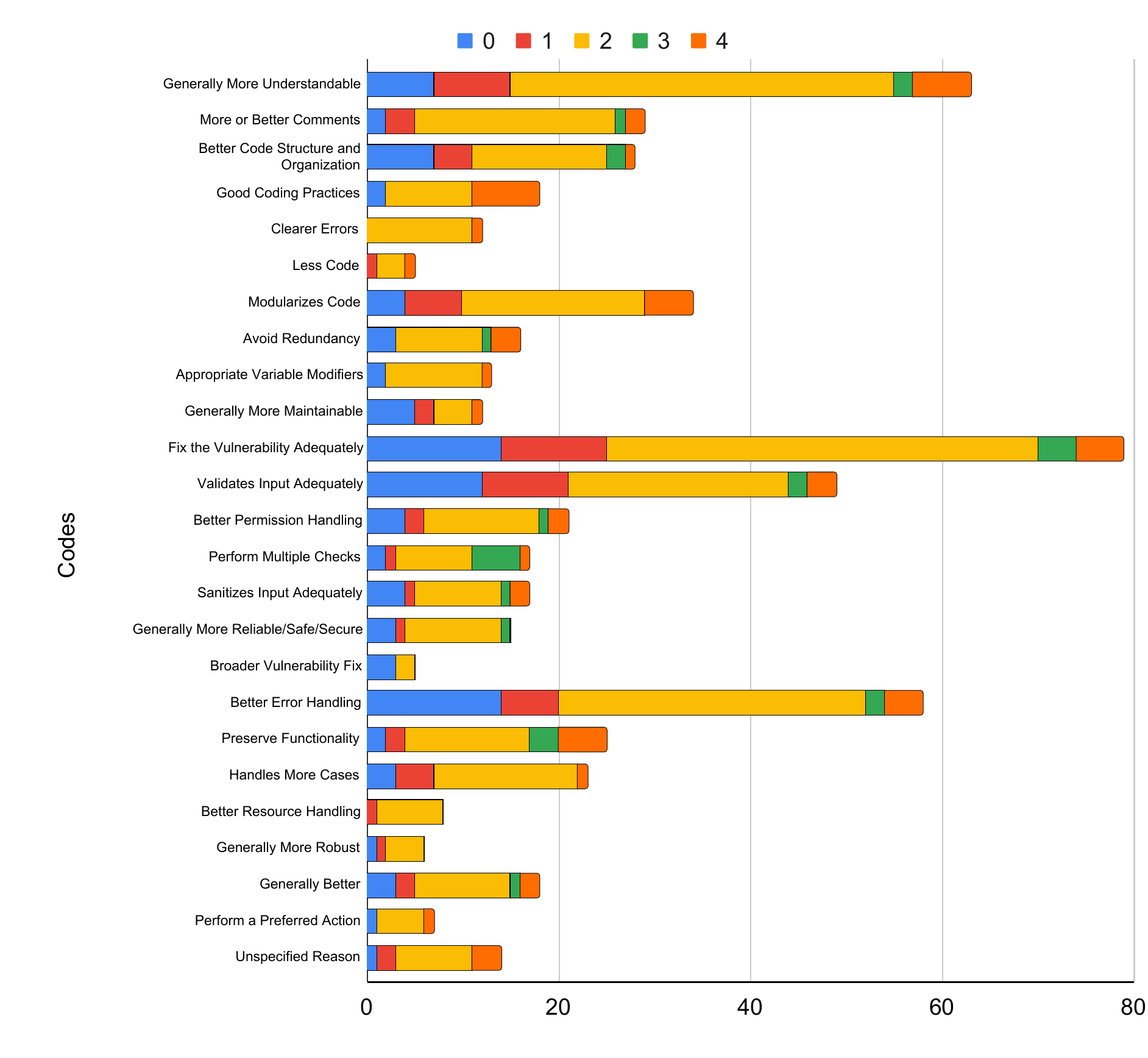}
        \caption{Experience with security vulnerabilities.}
        \label{fig:code_security}
    \end{subfigure}
    \caption{Number of answers labeled with the codes for each experience level across all answers (N=\allObservations).}
    \label{fig:code_experience}
\end{figure}

Figure~\ref{fig:code_java} shows the distribution of Java experience levels across individual codes.
Almost every code reflects the same distribution of participants across the five levels: most answers fell in level 2, followed by levels 1, 4, 3, and 0.

The 25 omnibus chi-square tests of independence revealed that only the code \codetheme{Broader Vulnerability Fix} yielded an adjusted p-value below 0.05, indicating that Java experience level is associated with participants' interest in patches that fix additional vulnerabilities or introduce extra security features.
The post-hoc analysis revealed that the standardized residual for experience level 0 exceeded the target threshold ($z_0 = 5.481 > 2.576$). 
The positive sign indicates that the code \codetheme{Broader Vulnerability Fix} appeared more frequently among participants with minimal Java experience than would be expected if we assumed the experience level had no influence.
This is also visible in Figure~\ref{fig:code_java}, where the share of answers of level 0 predominates more than what happens for other codes.
However, as this code was assigned by only five participants, the available evidence remains limited and should be interpreted with caution.
The results of the statistical tests and plots concerning the five themes can be found in the paper's online appendix~\cite{appendix}.

Figure~\ref{fig:code_security} shows the distribution of security experience levels for individual codes.
Almost every code reflects the same distribution of participants across the five levels: most answers fell in level 2, followed by levels 0, 1, 4, and 3.

As with the Java experience, we investigated whether the distribution of each code differed across security experience levels, using the same analysis procedure.
The codes \codetheme{Good Coding Practices} and \codetheme{Perform Multiple Checks} yielded adjusted p-values below 0.05, indicating that experience with security vulnerabilities is associated with participants' interest in patches that follow good coding practices and/or implementing layered checks (e.g., validating the input more than once).
We observed that the residuals at level 4 for code \codetheme{Good Coding Practices} ($z_4 = 3.950$) and at level 3 for code \codetheme{Perform Multiple Checks} ($z_3 = 4.498$) exceeded the threshold $2.576$.
Both residuals were positive, indicating that both codes appeared more frequently among participants with high security experience than would be expected if assuming the experience level had no influence.
This is also slightly visible in Figure~\ref{fig:code_security}, where the share of answers at the respective significant levels predominates more than what happens for other codes.
The results of the statistical tests and plots for the five themes are in the paper's online appendix~\cite{appendix}.

\rqanswer{2}{
Five main themes emerged from participants' free-text explanations for their patch selections: \codetheme{Security}, \codetheme{Code Comprehension}, \codetheme{Robustness and Functional Behavior}, \codetheme{Maintainability}, and \codetheme{Generic or Unspecified Reasons}, in descending order of frequency.
Almost half of the answers cited \codetheme{Security} as the primary reason for selecting a patch (e.g., a more adequate fix, stronger input validation, or better permission handling), followed by \codetheme{Code Comprehension}, with participants favoring readable patches and explanatory comments.

Professionals and academics had different priorities: professionals preferred patches that followed good coding practices (theme \codetheme{Code Comprehension} leads), whereas academics emphasized security improvements (theme \codetheme{Security} leads).
Participants with little Java experience favored patches that addressed multiple security flaws or introduced extra security features; participants with greater experience with security preferred patches that followed good coding practices or implemented layered security checks.
}

\subsection{Reaction on Attributed Provenance (\rqq{3})}
\label{subsec:rq3_results}

Table~\ref{tab:switches} summarizes decisions made by the two groups of participants (academics and professionals) and all participants together across the four rounds.
When the selected patch was attributed to a human developer, participants changed their choice in 12 of 186 cases (6.45\%), out of which 10 were from academics (6.33\%), and 2 were from professionals (7.14\%).
In both subgroups, the frequency of switches was similar.
In contrast, when the selected patch was attributed to an LLM, participants changed their choice almost 3 times as often, reaching 32 switches (17.20\%).
Of these, 30 were made by academics (18.99\%) and 2 by professionals (7.14\%).
The academics preserved this large difference between the two provenance sources (10 for human vs. 30 for LLM), whereas professionals showed no difference.
Unlike the previous case, the difference between the two subgroups is more evident, suggesting that academics tended to move away from LLMs, while professionals do not seem to think that the provenance makes any difference.

From the perspective of individual participants, each could perform up to 4 switches.
38 of the 93 ($\approx$40\%) participants switched their choice at least once: 34 of which were academics and 4 were professionals.
On average, a single participant switched 0.47 times, with 0.51 counting only academics (out of \acaValidParticipants) and just 0.29 if counting only professionals (out of \indParticipants).

To assess whether the number of switches between the two attributed provenance conditions differed significantly, we performed three chi-square tests of independence, one per contingency table.
Namely, whether there is statistical evidence that participants react differently when the selected patch was attributed to a human vs. to an LLM.
The test across all participants yielded an adjusted p-value of $0.004$ ($\chi^2=10.31$), confirming that the attributed provenance significantly affects the likelihood of changing the initial decision.
The corresponding odds ratio was $3.01$, indicating that participants were approximately three times more likely to switch their decision when the selected patch was attributed to an LLM rather than a human.
Similarly, the test for the academic group yielded an adjusted p-value of $0.002$ ($\chi^2=11.45$), with an odds ratio of $3.47$, suggesting an even stronger effect: academics were more than three times as likely to change their initial decision depending on the patch's attributed provenance.
Lastly, as expected, the test involving professionals was not rejected (adjusted $p=1.00$), indicating no significant evidence of a difference in switching behavior between the two provenance sources for this group.

\begin{table}[t]
\caption{Participants' reactions after learning the attributed provenance of the patch they selected across the four rounds. The participants either kept their original selection or switched to the other one.}
\label{tab:switches}
\resizebox{\linewidth}{!}{%
\begin{tabular}{p{3.5cm}|llr|llr|llr}
\hline
\multirow{2}{=}{\textbf{Attributed Provenance of the Selected Patch}} & \multicolumn{3}{c|}{\textbf{Academics (N=\acaObservations)}} & \multicolumn{3}{c|}{\textbf{Professionals (N=\indObservations)}} & \multicolumn{3}{c}{\textbf{All (N=\allObservations)}} \\
& \textbf{Keep} & \textbf{Switch} & \textbf{Total} & \textbf{Keep} & \textbf{Switch} & \textbf{Total} & \textbf{Keep} & \textbf{Switch} & \textbf{Total} \\
\hline
Human & 148 (93.67\%) & 10 (6.33\%) & 158 & 26 (92.86\%) & 2 (7.14\%) & 28 & 174 (93.55\%) & 12 (6.45\%)  & 186 \\
LLM & 128 (81.01\%) & 30 (18.99\%) & 158 & 26 (92.86\%) & 2 (7.14\%) & 28 & 154 (82.80\%) & 32 (17.20\%) & 186 \\
\hline
\textbf{Total} & \textbf{276 (87.34\%)} & \textbf{40 (12.66\%)} & \textbf{\acaObservations} & \textbf{52 (92.86\%)} & \textbf{4 (7.14\%)} & \textbf{\indObservations} & \textbf{328 (88.17\%)} & \textbf{44 (11.83\%)} & \textbf{\allObservations} \\
\hline
& \multicolumn{3}{l|}{$p=0.002$, $\chi^2=11.45$, $OR=3.47$} & \multicolumn{3}{l|}{$p=1.000$} & \multicolumn{3}{l}{$p=0.004$, $\chi^2=10.31$, $OR=3.01$}\\
\hline
\multicolumn{10}{l}{All p-values are adjusted with Bonferroni procedure.} \\
\end{tabular}
}
\end{table}

In summary, we observed that participants were significantly more inclined to reconsider their decision when the selected patch was attributed to an LLM.
Conversely, attributing the selected patch to a human developer substantially decreases participants' likelihood of switching to the other patch.
Even without counting professional participants, this effect remains evident.
Thus, we conclude that a patch's provenance can influence human judgment regardless of its technical characteristics.

\rqanswer{3}{
Participants were roughly three times as likely to change their initial decision when the selected patch was attributed to an LLM ($\approx$17\%) as when it was attributed to a human developer ($\approx$6\%).
This effect was driven almost entirely by academic participants: they switched decisions far more often for LLM-attributed patches ($\approx$19\%) than for human-attributed ones ($\approx$6\%).
In contrast, professionals switched at the same rate regardless of the attributed provenance ($\approx$7\% in both cases) and did so only a few times ($\approx$7\% of the total).
}
\subsection{Opinions on Attributed Provenance (\rqq{4})}
\label{subsec:rq4_results}

\subsubsection{Provenance Acknowledgment}

Table~\ref{tab:acks} reports the number of responses to question \ques[i]{R16} (see Table~\ref{tab:questions}) that explicitly acknowledged the provenance of the patches after this information was disclosed.
We observed that in more than 70\% of the cases, participants did not explicitly comment on the attributed provenance.
This indicates that for the majority of participants, provenance information was not a primary factor in their decision-making process.
Instead, among these non-acknowledging responses, participants either restated the motivation provided through \ques[i]{R14} (with minor rephrasing) or provided additional justification for their original choice.
These behaviors occurred in over 61\% and 37\% of the non-acknowledging cases, respectively.
In the remaining 3\% of cases, participants provided unclear explanations.

This trend of disregarding provenance is more pronounced among professionals, reaching almost 79\% (compared to over 72\% of academics), suggesting that professionals found this information even less valuable during their evaluation.
Among the professionals who did not acknowledge the attributed provenance, in almost four out of five cases (80\%) they simply restated their original motivations for the selection.
This behavior was significantly more common than among academics, where only 57\% of non-acknowledging participants restated their original motivations.

\begin{table}[t]
\small
\centering
\caption{Number of answers based on whether they explicitly acknowledged the provenance (`Ack'). Percentages are relative to the total number of answers (\textit{`Answers (Total)'}) of the respective group.}
\label{tab:acks}
\resizebox{\linewidth}{!}{%
\begin{tabular}{l|lll}
\toprule
& \textbf{Academics} & \textbf{Professionals} & \textbf{All} \\
\hline
\textbf{Answers (Total)} & \acaObservations & \indObservations & \allObservations \\
\hline
\textbf{Answers (NO Ack)} & 227 (71.84\%) & 44 (78.57\%) & 271 (72.85\%) \\
\textbullet{} \textbf{Restate Original Motivations} & 129 (56.83\%) & 35 (79.55\%) & 164 (60.52\%) \\
\textbullet{} \textbf{Provide Additional Motivations} & 93 (40.97\%) & 6 (13.64\%) & 99 (36.53\%) \\
\textbullet{} \textbf{Unclear/Unspecified Motivations} & 5 (2.20\%) & 3 (6.82\%) & 8 (2.95\%) \\
\hline
\textbf{Answers (YES Ack)$^*$} & 89 (28.17\%) & 12 (21.43\%) & \answersYesAck (27.15\%) \\
\bottomrule
\multicolumn{4}{l}{$^*$The distribution across themes and codes is reported in Table~\ref{tab:rq4_codes}.}
\end{tabular}
}
\end{table}

\subsubsection{The Recurring Codes and Themes}

For the remaining \answersYesAck responses that explicitly acknowledged the attributed provenance, we report the codes and themes derived from the thematic analysis in Table~\ref{tab:rq4_codes}.
The table also shows how many times each code and theme appeared in participants' answers, along with their relative frequencies across all \answersYesAck answers (called ``global frequency'') and within answers belonging to their theme (called ``in-theme frequency'').

\begin{table}[t]
\tiny
\caption{Codebook of the free answers to question \ques[i]{R16} ($i=1.2.3.4$) that explicitly acknowledged the provenance information (N=\answersYesAck) and commented on this, with the number of answers (\#) containing each code (white rows) and theme (gray rows). The column ``Freq.'' reports the global frequency (outside parentheses) and in-theme frequency (in parentheses).}
\label{tab:rq4_codes}
\resizebox{\linewidth}{!}{%
\begin{tabular}{p{1.6cm}p{4.2cm}|r>{\raggedleft\arraybackslash}p{0.8cm}|r>{\raggedleft\arraybackslash}p{0.8cm}|r>{\raggedleft\arraybackslash}p{0.8cm}}
\hline
\multirow{2}{=}{\textbf{Code/Theme Name}} &
\multirow{2}{=}{\textbf{Description (`The participant expressed...')}} & \multicolumn{2}{>{\centering\arraybackslash}p{1.45cm}|}{\textbf{Academics (N=89)}} & \multicolumn{2}{>{\centering\arraybackslash}p{1.45cm}|}{\textbf{Professionals (N=12)}} & \multicolumn{2}{>{\centering\arraybackslash}p{1.45cm}}{\textbf{All (N=\answersYesAck)}} \\
& & \textbf{\#} & \textbf{Freq.} & \textbf{\#} & \textbf{Freq.} & \textbf{\#} & \textbf{Freq.} \\ \hline

\rowcolor{gray20}
\textbf{Explicit Trust or Distrust} & Explicit trust or distrust toward the source of the selected patch. & \textbf{24} & \textbf{26.97\%}\par(\textbf{-}) & \textbf{1} & \textbf{8.33\%}\par(\textbf{-}) & \textbf{25} & \textbf{24.75\%}\par(\textbf{-}) \\ \cline{2-8}
Explicit Trust Toward Human & More trust, appreciation, or bias in favor of humans, even without objective motivation. & 20 & 22.47\% (83.33\%) & 1 & 8.33\% (100.00\%) & 21 & 20.79\% (84.00\%) \\ \cline{2-8}
Explicit Distrust Against LLM & Less trust, dislike, or bias toward LLMs, even without objective motivation. & 6 & 6.74\% (25.00\%) & 0 & 0.00\% (0.00\%) & 6 & 5.94\% (24.00\%) \\

\rowcolor{gray20}
\textbf{Human Capabilities} & Opinions concerning the capabilities of humans in producing security patches. & \textbf{32} & \textbf{35.96\%}\par(\textbf{-}) & \textbf{4} & \textbf{33.33\%}\par(\textbf{-}) & \textbf{36} & \textbf{35.64\%}\par(\textbf{-}) \\ \cline{2-8}
Humans Write Higher Quality Code & An opinion that humans write higher quality code (more understandable, natural, readable, reliable), at least compared to LLMs. & 2 & 2.25\% (6.25\%) & 2 & 16.67\% (50.00\%) & 4 & 3.96\% (11.11\%) \\ \cline{2-8}
Humans Have More Knowledge & An opinion that humans have more knowledge (e.g., of the domain, of the system), i.e., they know exactly where to change and have good reasons for making a change. & 14 & 15.73\% (43.75\%) & 1 & 8.33\% (25.00\%) & 15 & 14.85\% (41.67\%) \\ \cline{2-8}
Humans Have More Experience & An opinion that humans have more professional experience (e.g., in real-world code). & 6 & 6.74\% (18.75\%) & 0 & 0.00\% (0.00\%) & 6 & 5.94\% (16.67\%) \\ \cline{2-8}
Humans May Not Cover All Cases & An opinion that humans may not cover all cases in the code they write, at least not as much as LLMs. & 3 & 3.37\% (9.38\%) & 0 & 0.00\% (0.00\%) & 3 & 2.97\% (8.33\%) \\ \cline{2-8}
Humans Cover Edge Cases & An opinion that humans cover edge cases in the code they write, at least more accurately than LLMs. & 2 & 2.25\% (6.25\%) & 0 & 0.00\% (0.00\%) & 2 & 1.98\% (5.56\%) \\ \cline{2-8}
Humans Can Make Errors & An opinion that humans too can make errors. & 7 & 7.87\% (21.88\%) & 1 & 8.33\% (25.00\%) & 8 & 7.92\% (22.22\%) \\ \hline

\rowcolor{gray20}
\textbf{LLM Capabilities} & Opinions concerning the capabilities of LLMs in producing security patches. & \textbf{26} & \textbf{29.21\%}\par(\textbf{-}) & \textbf{4} & \textbf{33.33\%}\par(\textbf{-}) & \textbf{30} & \textbf{29.70\%}\par(\textbf{-}) \\ \cline{2-8}
LLMs Cover Many Cases & An opinion that LLMs are able to cover many cases in the code they generate, possibly even more than humans. & 7 & 7.87\% (26.92\%) & 0 & 0.00\% (0.00\%) & 7 & 6.93\% (23.33\%) \\ \cline{2-8}
LLMs Require Caution & Caution about the responses of LLMs (requiring double-checking) and how they were prompted. & 6 & 6.74\% (23.08\%) & 1 & 8.33\% (25.00\%) & 7 & 6.93\% (23.33\%) \\ \cline{2-8}
LLMs Lack Grounding & An opinion that LLMs often hallucinate, produce misleading information, or reproduce memorized data because they are trained to predict plausible text rather than grounded facts. & 6 & 6.74\% (23.08\%) & 1 & 8.33\% (25.00\%) & 7 & 6.93\% (23.33\%) \\ \cline{2-8}
LLMs May Generate Invalid Code & An opinion that LLMs can generate non-working code, e.g., not compilable or not suitable for the existing design. & 3 & 3.37\% (11.54\%) & 1 & 8.33\% (25.00\%) & 4 & 3.96\% (13.33\%) \\ \cline{2-8}
LLMs Struggle with Security & An opinion that LLMs struggle to handle security problems, which may result in generating vulnerable code or inadequate patches. & 4 & 4.49\% (15.38\%) & 0 & 0.00\% (0.00\%) & 4 & 3.96\% (13.33\%) \\ \cline{2-8}
LLMs Struggle with Complex Code & An opinion that LLMs struggle to handle long code or deal with complex aspects, e.g., arithmetic. & 2 & 2.25\% (7.69\%) & 1 & 8.33\% (25.00\%) & 3 & 2.97\% (10.00\%) \\ \hline
 
\rowcolor{gray20}
\textbf{Provenance Effect} & Opinions concerning the effect that knowing the provenance had on the decision process. & \textbf{37} & \textbf{41.57\%}\par(\textbf{-}) & \textbf{6} & \textbf{50.00\%}\par(\textbf{-}) & \textbf{43} & \textbf{42.57\%}\par(\textbf{-}) \\ \cline{2-8}
Provenance Does Not Affect Decision & Their stance that knowing the provenance did not affect their original decision. & 31 & 34.83\% (83.78\%) & 5 & 41.67\% (83.33\%) & 36 & 35.64\% (83.72\%) \\ \cline{2-8}
Human Provenance is an Added Value & Their stance that knowing the code originates from a human reinforces their decision. & 5 & 5.62\% (13.51\%) & 0 & 0.00\% (0.00\%) & 5 & 4.95\% (11.63\%) \\ \cline{2-8}
Recognizable Provenance by Code & That the provenance can be recognized by looking at the code's characteristics. & 2 & 2.25\% (5.41\%) & 1 & 8.33\% (16.67\%) & 3 & 2.97\% (6.98\%) \\
\hline
\end{tabular}
}
\end{table}

The most common theme across all participants was \codetheme{Provenance Effect}: almost half of the answers that acknowledged the attributed provenance (43\%) included comments regarding its impact on their decision process.
Specifically, 36\% reported that knowing the provenance did not affect the participant's original decision in the same round (\codetheme{Provenance Does Not Affect Decision}); this was also the most frequent code overall.
In these instances, participants claimed they evaluated the patches solely based on their technical quality, remaining unbiased by whether a human or an LLM generated the patch.
In a minority of cases (5\%), participants noted that discovering a human created their originally selected patch reinforced their decision, as humans add credibility.
On rare occasions (3\%), participants claimed they could infer the provenance by inspecting the code, having already guessed the source based on peculiar coding styles or comments indicative of an LLM.

The second most prevalent theme was \codetheme{Human Capabilities}: at least one out of three answers (36\%) expressed opinions concerning the capabilities of humans in writing security patches.
Most comments asserted that humans possess superior knowledge (e.g., regarding the specific domain or the system being patched); that is, they know precisely which part of the code requires modification and have sound reasoning for specific decisions (\codetheme{Humans Have More Knowledge}).
Similarly, some participants argued that humans have more experience working with real-world code (\codetheme{Humans Have More Experience}) and are capable of writing higher quality code (\codetheme{Humans Write Higher Quality Code}).
These observations implicitly suggest that participants believe LLMs lack such capabilities.
However, some participants also acknowledged human limitations; specifically, that humans can make errors (\codetheme{Humans Can Make Errors}) and may not cover all edge cases as comprehensively as an LLM (\codetheme{Humans May Not Cover All Cases}).
Despite these caveats, the general sentiment suggests that humans are perceived as having more positive than negative attributes in this context.

The third most prevalent theme was \codetheme{LLM Capabilities}, which acted as the dual of the previous theme, and appeared in almost one out of three answers (30\%).
Here, participants expressed opinions concerning the capabilities of LLMs in generating security patches.
No single code dominated the responses, and opinions varied widely.
For example, participants noted that LLMs often hallucinate, produce misleading information, or reproduce memorized data, as they learn to reproduce token sequences experienced during training rather than grounded facts (\codetheme{LLMs Lack Grounding}).
They also emphasized that processing LLM responses requires caution, such as double-checking outputs rather than accepting them as is (\codetheme{LLMs Require Caution}).
Conversely, there were comments regarding positive aspects of LLMs, such as their ability to cover many cases in the generated code, potentially even more than humans (\codetheme{LLMs Cover Many Cases}), which aligns with the code \codetheme{Humans May Not Cover All Cases}.
Nevertheless, the general trend suggests that LLMs are perceived to have more negative than positive aspects, contrary to the perception toward humans.

Lastly, the theme \codetheme{Explicit Trust or Distrust} emerged from one out of four answers (25\%).
Among these cases, the participant explicitly admitted greater trust toward humans (84\%) or distrust of LLMs (24\%).
These participants explicitly preferred humans once they learned the patch's attributed provenance, reflecting the phenomenon observed in the context of \rqq{3} (see Section~\ref{subsec:rq3_results}).

The distribution of codes and themes among academics aligns with the overall distribution, likely due to their significantly larger sample size compared to professional participants.
In this regard, identifying clear differences between the groups was challenging, as professionals accounted for only 12 acknowledging answers.
The only noteworthy distinction is that the \codetheme{Provenance Effect} theme was the most popular among professionals (50\% of their answers); in particular, they frequently stated that knowing the provenance did not affect their decision.

\subsubsection{Relationship with the Decision to Switch Patch}

Next, we examined the relationship between the acknowledgment of the attributed provenance and participants' decisions to keep or switch their originally selected patch.
The results are reported in Table~\ref{tab:ack_switches}.

\begin{table}[t]
\caption{Relation between participants' acknowledgment of provenance (rows) and the decision to keep or switch patches after learning the attributed provenance of patches (columns) across the four rounds.}
\label{tab:ack_switches}
\resizebox{\linewidth}{!}{%
\begin{tabular}{p{2.4cm}|llr|llr|llr}
\hline
\multirow{2}{=}{\textbf{Provenance Acknowledgment}} & \multicolumn{3}{c|}{\textbf{Academics (N=\acaObservations)}} & \multicolumn{3}{c|}{\textbf{Professionals (N=\indObservations)}} & \multicolumn{3}{c}{\textbf{All (N=\allObservations)}} \\
& \textbf{Keep} & \textbf{Switch} & \textbf{Total} & \textbf{Keep} & \textbf{Switch} & \textbf{Total} & \textbf{Keep} & \textbf{Switch} & \textbf{Total} \\
\hline
No & 204 (89.87\%) & 23 (10.13\%) & 227 & 42 (95.45\%) & 2 (4.55\%) & 44 & 246 (90.78\%) & 25 (9.22\%) & 271 \\
Yes & 72 (80.90\%) & 17 (19.10\%) & 89 & 10 (83.33\%) & 2 (16.67\%) & 12 & 82 (81.19\%) & 19 (18.81\%) & 101 \\
\hline
\textbf{Total} & \textbf{276 (87.34\%)} & \textbf{40 (12.66\%)} & \textbf{\acaObservations} & \textbf{52 (92.86\%)} & \textbf{4 (7.14\%)} & \textbf{\indObservations} & \textbf{328 (88.17\%)} & \textbf{44 (11.83\%)} & \textbf{\allObservations} \\
\hline
& \multicolumn{3}{l|}{$p=0.046$, $\chi^2=4.652$, $OR=2.09$} & \multicolumn{3}{l|}{$p=0.148$, $\chi^2=2.089$, $OR=4.20$} & \multicolumn{3}{l}{$p=0.033$, $\chi^2=6.484$, $OR=2.28$}\\
\hline
\multicolumn{10}{l}{All p-values are adjusted with Benjamini-Hochberg procedure.} \\
\end{tabular}
}
\end{table}

Among the \answersNoAck cases, participants retained their selection in 246 cases (91\%); conversely, in the remaining 25 cases (9\%), they switched patches.
In contrast, among the \answersYesAck cases, participants retained their selection in 82 cases (81\%); thus, in the remaining 19 cases (19\%), they switched patches.

Contrary to expectations, even when participants did not acknowledge the attributed provenance, they still switched patches in nearly one out of ten cases.
Although we lack explicit confirmation regarding the cause (as participants did not always disclose this), this may have occurred because participants reconsidered their choice when given the opportunity to re-inspect the patches.
Indeed, in at least 30\% of cases, participants provided additional motivations that suggested there was a \textit{second thought}.

However, comparing the proportions between the two groups (acknowledgment `No' vs. `Yes') reveals a \textit{higher tendency to switch when provenance is acknowledged}. We assessed whether the number of switches differed significantly between the two acknowledgment conditions via three chi-square tests of independence.
Specifically, we tested whether participants who acknowledged provenance switched patches significantly more often than those who did not.

The test across all participants yielded an adjusted p-value of $0.033 < 0.05$ ($\chi^2 = 6.484$), confirming that participants were more likely to switch patches when they explicitly acknowledged provenance.
In particular, the odds of switching increased by a factor of 2.28.
An example was answer \#137 (\answerVerbatim{I am not sure to what extent the LLM has been prompted to fulfill the needs for the specific patch, and as a non expert user I am biased to let the human expert patch through in the absence of this clarifying information}).
In this instance, the participant preferred trusting the human expert because they were uncertain how the LLM was prompted (i.e., whether the LLM knew exactly what it was doing), leading them to change their initial selection.
The codes assigned to this case were \codetheme{LLMs Require Caution} and \codetheme{Explicit Trust Toward Human}.
The significance also holds in the academic group ($p=0.046 < 0.05$, $OR = 2.09$) but not for professionals ($p=0.148>0.05$).

Finally, we verified whether participants who stated that provenance had no effect on their decision process (i.e., were assigned code \codetheme{Provenance Does Not Affect Decision}) retained their original selection, as expected.
We confirmed that none of these participants changed their minds, aligning with their disclosed responses.

\rqanswer{4}{
After the attributed provenance is disclosed, over 70\% of participants' free-text responses did not explicitly acknowledge it in their explanations, with professionals disregarding it even more frequently than academics.
Among those who did acknowledge it, four main themes emerged: \codetheme{Provenance Effect}, \codetheme{Human Capabilities}, \codetheme{LLM Capabilities}, and \codetheme{Explicit Trust or Distrust}, in descending order of frequency.
Almost half of these answers ($\approx43\%$) concerned the effect that knowing the provenance can have on the selection process, predominantly stating that this information does not influence their decisions (indeed, this group never changed their mind).
However, other responses admitted that bias does take place, especially in favor of humans and against LLMs.
Notably, participants who explicitly acknowledged provenance were at least two times as likely to change their initially selected patch.
}

\section{Discussion}
\label{sec:discussion}

\textbf{Characterizing Human Bias.}
In the analysis of \rqq{3}, we observed that once participants had the opportunity to move away from a patch they learned was GenAI-made, they tended to prefer the one attributed to a human.
This was clearly visible among academic participants, but not among professionals (due to the limited sample size).
In the analysis of \rqq{4} we also found participants explicitly expressing bias in favor of humans and also bias against LLMs.
Together, these findings suggest the presence of a \textit{provenance-related bias}.
One possible explanation is participant expertise.
Most participants in our pool were Master's students, who, on average, are less experienced than professionals; this might translate into less skill in critically reviewing a patch, making them more inclined to trust a human than a GenAI model (or perhaps an automated tool in general, though this was an aspect left outside our experiment).
Conversely, participants with more experience might feel confident enough in their technical judgment to be less influenced by a patch's source.
Follow-up studies are needed to explore whether experience is actually a key driver for this bias.
Additionally, a replication of this experiment where the specific GenAI model name is disclosed could help characterize the bias to verify whether an authoritative name like GPT or Claude gives more (or less) trust than a generic AI or LLM.
Indeed, we hypothesize that patches from older or less reputable LLMs would be less likely to be preferred than patches from frontier models.
Lastly, we cannot rule out the possibility that participants reconsidered the patch selected because they had the opportunity to re-inspect it and had a second thought rather than because they did not trust the attributed provenance. 
In \rqq{4} we observed that participants often did not comment on provenance but still changed their selection, which may suggest they looked more closely at the patches and changed their mind.
Unfortunately, we do not have sufficient evidence for this, as the answers do not explicitly state that the switch was due to closer inspection.

\textbf{Further Experiments with Humans.}
In our experiment, participants were shown two patches per round and asked to indicate which one was better.
As they could only choose between the two given options, their selection may not necessarily reflect the patch they would have produced or used themselves in a similar situation.
This is suggested by the analysis of \rqq{2}, in which some open-ended answers hint that participants are not fully convinced by the selected option.
To explore this, future investigations could use a different experimental design in which participants are given candidate patches but are free to modify them, or even write a completely different patch from scratch.
In this setting, we could measure the extent of the changes made as a proxy for effort and compare how quickly participants produce a working fix relative to a control group that receives no candidate patches.
This design would also let us assess the \textit{anchoring bias}, i.e., whether the patches participants eventually produce closely resemble the initial patches they were shown, and whether this anchoring affects the quality of the final patch.
To test this specifically, some participants would need to be given intentionally incorrect, but still plausible, patches as a starting point, allowing us to check whether they carry over the errors.
Beyond looking at individual patches, we also need to study how humans interact with complete AVR tools, rather than just patches.
This calls for an experiment in which participants use real tools, whether GenAI-powered or not, to repair real vulnerabilities, measuring the extent to which these tools help or hinder them.

\textbf{Recommendations for AVR Tool Designers.}
Based on our experimental results, we derived a set of recommendations for AVR tool designers.
First, AVR tools should optimize for patch \textit{understandability and maintainability}, alongside security correctness and the avoidance of functionality regressions; these last two aspects remain essential and should not be traded off.
Tools should also penalize candidate patches that contain code smells or violate clean coding principles. For example, good patches keep sanitization logic in separate methods and raise fine-grained exceptions rather than generic ones.
Second, AVR tools should generate patches with \textit{more code comments}, provided these comments are not redundant.
In particular, comments should explain how the patch prevents the vulnerability rather than simply describing what each line of code does.
This facilitates the assessment of a patch's correctness.
Alternatively, tools could produce a separate, detailed explanation to accompany the patch.
Third, AVR tools could make \textit{patch installation more straightforward} by applying the patch to the codebase automatically, rather than requiring developers to copy and paste the changes by hand, a manual process that is prone to introducing errors.

\section{Threats to Validity}
\label{sec:ttv}

\subsection{Internal Validity}

\textbf{Experimental Design.}
Participants may have had a systematic bias toward the first patch their eyes fell on (e.g., the left one) or realized that the provenance followed a fixed set of rules, making the disclosure of the real origin seem unbelievable and leading them to intentionally align (or go against) the experimenters' hypotheses.
The crossover-balanced design~\cite{massacci2024addressing} of Table~\ref{tab:design} addresses both: the two experiment factors \llmForFirstPatch and \selectedIsHuman alternate across the four rounds within each group, so no participant sees the human attribution consistently attached to the same position, the same LLM, or their own selection.
Moreover, the disclosure was presented on a separate page after the participant made their selection and submitted their justification, to discourage them from going back and revising their initial decision.
All participants were debriefed at the end of the questionnaire, which revealed the true origin of all patches for transparency.

\textbf{Fatigue.}
Each round required reading a vulnerability description and its corresponding source code, inspecting two patches, answering 12 Likert-scale items, making two selections, and writing two free-text motivations (albeit optional).
Repeating this four times may induce fatigue, causing the answers in later rounds to be less focused than those in the first rounds.
Studies~\cite{ackerman2009test,daun2017controlled} have shown that increased test length induces cognitive fatigue that affects performance, and that long tasks in (student) experiments increase dropout rates and disengagement.
Daun et al.~\cite{daun2017controlled} surveyed 163 student experiments and found that only 21\% of them were designed to last less than two hours.
The questionnaire was designed to be solvable in 75 minutes, so we set a 75-minute cap.
We further limited fatigue risk by limiting the rounds to four, selecting vulnerabilities localized to a single file, and running the comprehensive video tutorial beforehand so participants did not have to learn the questionnaire while completing it.
On average, the participants who passed the right number of attention checks submitted the questionnaire within 68 minutes.

\textbf{Learning Effect.}
The four vulnerabilities were presented in the same order, so the round number is confounded with vulnerability, making it hard to separate the learning effect from the vulnerabilities' specific characteristics.
The presence of the confounding does not invalidate the study's results, as all statistical analyses are based on aggregating observations per round and the factors are perfectly balanced across rounds (see Table~\ref{tab:design}).
Therefore, any learning effect will affect all levels of treatment equally.
In addition, we used four vulnerabilities across three types of weaknesses and three projects.
Thus, transferring knowledge from one round to another round for vulnerability-specific features is highly unlikely.
Participants could at most gain procedural experience that made them faster at completing the later round.

\textbf{Recruitment of Participants.}
All participants joined voluntarily.
Students were encouraged through bonus points in the respective study courses.
Such an offer might attract students motivated by the bonus points rather than by the task itself.
A participant was granted bonus points only if they completed the task with serious involvement, regardless of the decisions they made in the questionnaire or whether they consented to the use of their responses for research purposes (they had the option to opt out without consequences, as some participants did).
Thus, there was no bias towards any specific response by student participants.

We received 136 complete responses in total.
As described in Section~\ref{subsec:data_analysis}, we excluded 17 responses from participants who did not authorize the processing of their data and 26 responses because they failed the majority of attention checks.
All excluded responses were from academic participants; none of the professional participants' responses were discarded.
The threshold value is deliberately low because errors can occur even among participants with serious involvement.

\textbf{Data Leakage During Patch Generation.}
Each of the subject vulnerabilities (and their fixes) is dated to 2018 and recorded in well-known databases.
Hence, it is possible that \gpt and/or \claude encountered them during their training.
Zhou et al.~\cite{zhou2025lessleak} conducted a large-scale analysis of 83 software engineering benchmarks and found that data leakage in software engineering tasks is minimal (2.8\% in Java).
Additionally, the models may have already forgotten the patches due to catastrophic forgetting~\cite{zhai2024investigating}.
Nevertheless, we manually inspected the generated patches (see Section~\ref{subsec:preparation}) and confirmed they were sufficiently different from the developer patches.

\subsection{External Validity}

\textbf{Participant Representativeness.}
Most of our participants were Master's students, which raises the question of how far (or whether) the results generalize to a real-world setting of developers.
Naiakshina et al.~\cite{naiakshina2020onconducting} replicated a password-storage study with computer science students, freelancers, and company developers, and found that students are a valid proxy for professional developers in security-related developer studies.
Our participants are therefore best interpreted as proxies for early-career developers, which is a meaningful population for our purposes, as they are the most likely adopters of AI-based repair assistance.
The professional sample provides a complementary perspective, but it is quite small, drawn from three companies, and skewed toward high experience (10 out of 14 reported the maximum level in both Java programming and security vulnerabilities).
We believe the findings about professionals are indicative rather than generalizable.

\textbf{Review Scenario.}
Both patches from each round were plausible and were manually validated beforehand to confirm that they adequately fixed the vulnerability, so participants chose between two appropriate alternatives.
Real AVR tools might produce incorrect and overfitted patches~\cite{Bui:emse2023:apr4vul}.
Therefore, our study characterizes the preference among good candidates, not the ability to spot bad ones, which was the focus of Papotti et al.~\cite{papotti2024acceptance}.
In addition, participants could choose exactly one preference per round, without the option of declaring the two patches equivalent or both inadequate.
They could neither modify the patches nor execute the unit tests nor browse the codebase for context.
These simplifications keep the task comparable across participants but remove aspects of a realistic code review workflow.
This aspect, however, was outside the scope of this study.

\textbf{Selected Vulnerabilities and Patches.}
The current experiment used four real-life vulnerabilities from three open-source Java projects, covering three types of weaknesses.
The patches were generated by two commercial LLMs in a realistic zero-shot setting.
Our results may not be generalizable to memory-unsafe languages such as C/C++ or to more complex multi-file vulnerabilities.
These vulnerabilities were omitted to keep the sessions focused and simple.
Likewise, newer AVR approaches (e.g., using agentic AI) may generate patches with different characteristics, which could influence the selection criteria.
A small number of objects is inherent to controlled experiments with humans, where the task duration is important; we mitigated this by maximizing the diversity of weakness types among the selected vulnerabilities.

\subsection{Construct Validity}

\textbf{Absence of Human-authored Patches.}
All the patches shown in the experiment were generated by LLMs, so there is a risk that the study measures a preference between \gpt and \claude rather than between AI-generated and human-authored patches.
This was an intentional design choice, and this does not affect our research questions.
For \rqq{1} and \rqq{2}, provenance plays no role.
When making the first selection, participants were told only that both patches were plausible, and the names of the models were never disclosed at any point during the experiment.
For \rqq{3} and \rqq{4}, we investigated the effect of the \textit{attributed} provenance, and keeping the two code samples unchanged before and after the disclosure is what allows us to assign the re-selection to the label alone.
Showing an actual human patch (e.g., the historical fix) could possibly have confounded provenance with differences in code characteristics, as happens in other studies~\cite{alarcon2020would,ryan2019trust}.
Hence, we measure a bias toward the idea of human authorship rather than toward actual human patches, and we phrase our conclusions accordingly.

\textbf{Qualitative Coding.}
The qualitative coding process was primarily conducted by one author, supported by \deepseekLong to generate preliminary codes for approximately two-thirds of the answers.
All the codes assigned by the LLM were reviewed by the inspector, and errors were corrected.
To ensure accuracy, a random sample of 100 answers ($\approx$27\% of the dataset) across research questions \rqq{2} and \rqq{4} underwent a validity check by another co-author, who verified the validity of the codes assigned to each answer (labeled as YES/NO).
The initial agreement rates were high (92\% and 96\% for \rqq{2} and \rqq{4}, respectively), supporting the validity of the main inspector's coding.
Any disagreements identified during the validation were discussed and resolved through consensus.
The review and consensus process mitigates the risk of individual bias, though we acknowledge that independent double-coding of the entire dataset would offer higher reliability, 
Nevertheless, the analysis of the disagreed cases resulted in changing the codes assigned to answers without altering the underlying codebook content.

\textbf{Patch Correctness.}
Patches were automatically evaluated using Proof-of-Vulnerability test cases, which confirm the correctness of the fixes but cannot guarantee coverage of future security regressions.
This was unnecessary for the experiment, as the participants were only asked to assess the patch's suitability in that specific scenario.
Nevertheless, we also made a manual inspection of the plausible patches to further support their suitability.

\subsection{Conclusion Validity}

\textbf{LLMs as Confounding Factors.}
When we fit the logistic regression models for \rqq{1}, we asked whether the LLM that produced a certain patch could indirectly affect how participants judged the patches (across the six aspects), even though participants never knew at any point the exact LLM behind each patch.
Indeed, in \rqq{1} our goal was not to determine which LLM was preferred over another, but only whether the six aspects affected the selection.
Thus, we checked the average patch judgments across the four rounds by the LLM that generated them, as reported in Table~\ref{tab:judgments}.
Among all observations (\allObservations), the largest mean differences were observed for direct applicability ($Dir$), with \claude's patches scoring 0.18 higher than \gpt's (3.67 vs. 3.49), followed by preservation of functionality ($Prs$, 0.15; 3.85 vs. 3.70) and vulnerability removal ($Rem$, 0.14; 3.90 vs. 3.76).
The same direction held for academic participants, although the differences were smaller.
Among professionals, \claude's patches scored higher on all six aspects; the largest differences concerned direct applicability (0.25), vulnerability removal (0.23), and preservation of functionality (0.20).
Nevertheless, the differences were modest overall.
Since Table~\ref{tab:judgments} reports descriptive averages, these values alone do not establish whether the generating LLM significantly affected participants' judgments.

\begin{table}[t]
\caption{Average scores (from 1 to 5) that participants assigned to the six aspects of the patches made by \gpt and \claude across the four rounds.}
\label{tab:judgments}
\resizebox{\linewidth}{!}{%
\begin{tabular}{l|cc|cc|cc}
\hline
\multirow{2}{*}{\textbf{Aspect}} & \multicolumn{2}{c|}{\textbf{Academics (N=\acaObservations)}} & \multicolumn{2}{c|}{\textbf{Professionals (N=\indObservations)}} & \multicolumn{2}{c}{\textbf{All (N=\allObservations)}} \\
& \textbf{\gptShort} & \textbf{\claudeShort} & \textbf{\gptShort} & \textbf{\claudeShort} & \textbf{\gptShort} & \textbf{\claudeShort} \\
\hline
Understand intention ($Int$) & 3.94 & 3.95 & 4.16 & 4.25 & 3.98 & 4.00 \\
Understand modified behavior ($Mod$) & 3.80 & 3.88 & 4.14 & 4.20 & 3.85 & 3.93 \\
Understand vulnerability fix ($Fix$) & 3.70 & 3.79 & 4.14 & 4.25 & 3.77 & 3.86 \\
Believe preserved functionality ($Prs$) & 3.68 & 3.82 & 3.84 & 4.04 & 3.70 & 3.85 \\
Believe vulnerability removed ($Rem$) & 3.72 & 3.84 & 4.02 & 4.25 & 3.76 & 3.90 \\
Believe direct applicability ($Dir$) & 3.44 & 3.60 & 3.77 & 4.02 & 3.49 & 3.67 \\
\hline
\end{tabular}
}
\end{table}

To complement this, we also analyzed the general selection trends across rounds.
Table~\ref{tab:rounds} presents the frequency with which patches from the two LLMs (\gpt or \claude) were selected in each round prior to provenance disclosure (i.e., when participants are informed that the patch they selected was attributed to a human or an LLM), as well as the variations resulting from re-selection after disclosure.
We observe that participants selected \claude's patches more frequently than \gpt's.
We remark that participants \textit{were unaware} of the exact LLM source behind these patches; thus, these results do not reflect participant preference for one model over another, but rather merely reflect the outcome of their selection.
This prevents us from concluding that \claude produces better patches than \gpt, but suggests at most that \claude tends to produce patches that participants prefer.
In some cases (44 out of \allObservations), participants reconsidered their original selection (more details in \rqq{3}).
Nevertheless, the overall balance from the LLMs' perspective remains relatively stable, supporting the fact that the specific model did not add noise to the re-selection process.

\begin{table}[t]
\caption{Number of times the patch of an LLM (\gpt and \claude) was selected in each round before the provenance (human or LLM) is disclosed. The number in parentheses reports the variation after re-selection following disclosure.}
\label{tab:rounds}
\resizebox{\linewidth}{!}{%
\begin{tabular}{l|ll|ll|ll}
\hline
\multirow{2}{*}{\textbf{}} & \multicolumn{2}{c|}{\textbf{Academics (N=\acaObservations)}} & \multicolumn{2}{c|}{\textbf{Professionals (N=\indObservations)}} & \multicolumn{2}{c}{\textbf{All (N=\allObservations)}} \\
& \textbf{\gptShort} & \textbf{\claudeShort} & \textbf{\gptShort} & \textbf{\claudeShort} & \textbf{\gptShort} & \textbf{\claudeShort} \\
\hline
Round \#1 & 26 ($-$1) & 53 ($+$1) & 7 ($/$) & 7 ($/$) & 33 ($-$1) & 60 ($+$1) \\
Round \#2 & 33 ($+$2) & 46 ($-$2) & 4 ($+$1) & 10 ($-$1) & 37 ($+$3) & 56 ($-$3) \\
Round \#3 & 26 ($+$3) & 53 ($-$3) & 3 ($/$) & 11 ($/$) & 29 ($+$3) & 64 ($-$3) \\
Round \#4 & 31 ($-$4) & 48 ($+$4) & 3 ($+$1) & 11 ($-$1) & 34 ($-$3) & 59 ($+$3) \\
\hline
\textbf{Total} & \textbf{116 ($/$)} & \textbf{200 ($/$)} & \textbf{17 ($+$2)} & \textbf{39 ($-$2)} & \textbf{133 ($+$2)} & \textbf{239 ($-$2)} \\
\end{tabular}
}
\end{table}

\textbf{Statistical Testing.}
For categorical variables, we used Pearson's chi-squared test of independence as the expected cell counts were sufficiently large to justify this test.
When we ran multiple tests within the same family, e.g., comparing codes and themes between academics and professionals, we adjusted the resulting p-values to control the false discovery rate using the Benjamini-Hochberg procedure~\cite{bh:correction}.
For \rqq{3}, we applied a Bonferroni correction instead.
For the code and theme comparisons, the Benjamini-Hochberg procedure was chosen over more conservative correction procedures (e.g., Bonferroni) to maintain statistical power while reducing false positives.
We analyzed only the adjusted p-values and reported only the differences that survived the correction.
To contextualize the results of significant tests, we also calculated Odds Ratios (OR) to measure the strength of the association.

\section{Related Work}
\label{sec:rw}

\subsection{Technical Assessment of Security Patches}

The quality assessment of bug-fixing patches has been investigated from multiple perspectives, with a strong focus on correctness, validation, and ranking \cite{zhang2023survey}. In Automated Program Repair (APR), patch correctness is commonly approximated through dedicated test suites, such as JUnit test cases, which exercise the target functionality and serve as practical oracles for validating candidate patches. To this end, execution-based metrics, including patch plausibility and correctness, are commonly used to assess whether generated patches compile, pass the available tests, and satisfy the expected behavior captured by the test suite \cite{gazzola2018automatic}. In learning-based repair settings, match-based metrics such as accuracy and BLEU have also been used to compare generated patches against human-written reference fixes \cite{zhang2023survey}. However, these metrics provide only a partial view of patch quality, since patches that pass the available tests may still be semantically incorrect, overfit the test suite, or introduce unintended side effects \cite{liu2021critical,fei2025patch}. Likewise, syntactic similarity to a reference patch does not necessarily imply correctness. These limitations become particularly relevant in Automated Vulnerability Repair (AVR), where the expected patch must not only preserve functional correctness but also eliminate the underlying security weakness without introducing new vulnerabilities or weakening existing security assumptions \cite{hu:usenix2025:avr:sok}.

The validation of AVR-generated patches has been approached through both static and dynamic methods, depending on whether the patched program needs to be executed.
Static validation methods inspect the patched code without execution, for instance by using resource-leak detectors, contract analysis tools, or other program analysis techniques to identify residual defects and security-relevant inconsistencies~\cite{hu:usenix2025:avr:sok,li:usenix2025:sok,Utture:fse2023}.
Dynamic validation methods, in contrast, execute the patched program against test inputs, regression test suites, Proof-of-Vulnerability (PoV) tests, or fuzzing campaigns to assess whether the vulnerability has been removed and whether the patch preserves the expected functionality~\cite{Zhang:issta22,Bui:emse2023:apr4vul,wang:acl25:cvebench,mei:eurosp26:arvo,lee:neurips26:secbench}.
Several benchmark datasets have emerged in recent years to support the empirical assessment of AVR tools \cite{bui2022vul4j,wang:acl25:cvebench,mei:eurosp26:arvo,lee:neurips26:secbench,wei:2025:patcheval,wang:2025:vulnrepaireval}. At their best, such benchmarks provide real-world or synthetic vulnerabilities together with one or more exploits, corresponding patches, and vulnerability reports~\cite{hu:usenix2025:avr:sok}. Nevertheless, and despite their importance, not all datasets contain executable exploits, as these can be difficult to reproduce for certain vulnerability types or software environments. This limits the extent to which AVR-generated patches can be evaluated against concrete attack scenarios and, in turn, constrains the development and comparison of repair techniques for security bugs. Altogether, these limitations suggest that technical validation alone may be insufficient to establish whether AVR-generated patches are trustworthy in practice.

\subsection{Developer's Views on Automatic Vulnerability Patching}

Alongside the technical benchmarking of APR and AVR techniques, a growing body of research has examined the usability and trustworthiness of candidate patches \cite{noller2022trust}. This line of work focuses on the factors that influence developers' willingness to adopt automatically generated fixes, including their perceptions of patch usefulness \cite{winter2022developers}, understandability \cite{eladawy2024automated}, and maintainability \cite{fry2012human,steenhoek2025closing}. Ryan et al. \cite{ryan2019trust} conducted a series of controlled experiments to investigate how developer expertise shapes trust in APR-generated patches. They found that novice programmers were more willing to adopt automatically generated fixes, whereas expert programmers tended to prefer human-written ones, which they perceived as more trustworthy. Similarly, a survey study by Noller et al. \cite{noller2022trust} showed that trust in APR fixes can be substantially increased when patches are accompanied by explanations and test reports produced by static and dynamic analysis tools. In the same vein, Winter et al. \cite{winter2022developers} observed that developers prefer to remain in the patch generation loop by selecting and validating fixes themselves, rather than relying on fully automated program repair processes. However, while APR tools may increase debugging efficiency, they may also induce automation bias, leading developers to accept generated patches with insufficient scrutiny despite potential quality concerns \cite{eladawy2024automated}.

Despite growing efforts to develop reliable AVR techniques, the factors that shape their perceived trustworthiness and acceptance among software practitioners remain underexplored. Papotti et al. \cite{papotti2024acceptance} conducted one of the few studies in this area, examining developers' acceptance of potentially incorrect security patches. In a two-stage controlled experiment, they observed that participants were able to identify incorrect patches more easily than correct ones and also that correct patches were not confused with partially correct ones.
Nevertheless, participants were also more inclined to adopt patches generated by security-specific AVR tools than those produced by generic APR tools, even when the patches were incorrect. As AI-based techniques become increasingly prominent in AVR, recent work has begun to examine developers' attitudes and perceptions toward AI-powered AVR tools. For instance, Steenhoek et al. \cite{steenhoek2025closing} found that adoption and trust can be hindered by the complexity of generated fixes and the lack of explainability. Still, practitioners' trust in AI-generated vulnerability patches remains insufficiently understood, particularly from a decision-making perspective. More specifically, limited attention has been paid to the patch quality indicators practitioners rely on when assessing such fixes, and to how their attitudes toward AI technologies shape these assessments.

\section{Conclusion}
\label{sec:conclusion}

The experiment results showed that the human judgments on patch aspects like (i) the extent to which it \textit{preserves existing functionality}, (ii) how well they \textit{understood its security fix}, (iii) the extent to which they believed it \textit{removes the vulnerability}, and (iv) whether it can be \textit{applied without further changes} are the main drivers for patch selection.
This suggests that patches that are \textbf{effective}, \textbf{clear}, \textbf{do not introduce bugs}, and \textbf{easy to install} are preferred.

Among the motivations participants provided to explain their patch selection process (\rqq{2}), they reported that patch correctness (\textit{how well the patch fixed the vulnerability}) and comprehensibility (\textit{how easily the patch can be understood}) are the primary reasons.
Academics placed more emphasis on the first, while professionals paid more attention to the latter; however, this correlates with experience (i.e., experienced participants are more interested in comprehensibility than correctness).

Knowing the provenance of patches (\rqq{3}) influences human judgment regardless of their technical characteristics.
Participants were roughly three times as likely to change their initial decision when the selected patch was attributed to an LLM.
This phenomenon was not observed among professionals, though the limited sample size did not allow a strong conclusion.

Among the opinions participants provided to comment on the attributed provenance (\rqq{4}), over 70\% did not explicitly acknowledge the disclosed provenance, with professionals disregarding it even more frequently than academics.
Participants who explicitly acknowledged provenance were at least twice as likely to revise their initially selected patch.
In some cases, the participants admitted bias in favor of humans and against LLMs; combined with \rqq{3}, this further supports the existence of \textbf{provenance-related bias}.

In this paper, we focused on human trust in AI-generated patches for security vulnerabilities.
Nevertheless, the results of this experiment can be leveraged to further study human attitude toward AI-driven AVR tools in general.
One option is to employ the Technology Acceptance Model (TAM) to assess the \textit{perceived usefulness} and \textit{ease of use} of AVR tools, particularly those driven by LLMs and AI more broadly.
This would require a different experimental design, in which participants are given access to AVR tools to fix vulnerabilities themselves.
From a different angle, the insights gathered from industry professionals could be expanded through semi-structured interviews and focus groups to better understand exactly which properties of AI-generated patches are most desirable for increasing human trust in them.

\begin{acknowledgements}
This work was partially supported by EU-funded project Sec4AI4Sec (grant no. 101120393).
\end{acknowledgements}

\section*{Conflict of interest}
The authors declare that they have no conflict of interest.

\section*{CRediT author statement}

\textbf{Emanuele Iannone}: Conceptualization, Methodology, Formal analysis, Investigation, Data Curation, Writing - Original Draft, Writing - Review \& Editing, Visualization; 
\textbf{Maria Camporese}: Methodology, Formal analysis, Investigation, Data Curation, Writing - Original Draft, Writing - Review \& Editing, Visualization; 
\textbf{Nicolás E. Díaz Ferreyra}: Conceptualization, Methodology, Formal analysis, Investigation, Writing - Original Draft;
\textbf{Gábor Antal}: Conceptualization, Software, Resources, Writing - Review \& Editing;
\textbf{Péter Hegedűs}: Supervision, Writing - Review \& Editing;
\textbf{Fabio Massacci}: Conceptualization, Methodology, Writing - Review \& Editing, Supervision, Funding acquisition; 
\textbf{Riccardo Scandariato}: Conceptualization, Methodology, Writing - Review \& Editing, Supervision, Project administration, Funding acquisition.

\section*{Data Availability}
\label{sec:data:availability}
The supplementary material for this paper is available on \textsc{Zenodo} and includes participant responses (raw and processed), the outcome of the thematic analysis of free-text answers, the scripts and spreadsheets to reproduce the analyses to answer the research questions, and a copy of the experimental material~\cite{appendix}.

\bibliographystyle{spmpsci}
\bibliography{biblio}

\newpage
\appendix
\section{\deepseekShort's Prompt for the Code Assignment (\rqq{2})}
\label{appendix:prompt_rq2}

The following prompt was given to \deepseekLong through the \textsc{ChatAI} service of \textsc{Academic Cloud} (\url{https://chat-ai.academiccloud.de/}) on June 10, 2026.
This service hosts several models, including \deepseekLong, in Germany and guarantees that the data is processed in loco, securely stored, and not used for training other models.
The model was queried with a \textit{temperature} of 0.5, \textit{top\_p} of 0.5, and a ``low'' reasoning setting.

\prompt{}{
You are an expert in thematic analysis and in software security. Your goal is to perform qualitative open coding on a set of participant responses. These responses come from a questionnaire administered during an experiment in which participants were asked to select one of two patches that fixed a certain vulnerability (the one believed to be correct). The responses you will review are the motivations they disclosed to explain their choice.

\vspace{\baselineskip}
\noindent
You will be given:
\begin{enumerate}[leftmargin=*,label=\arabic*.]
    \item A Codebook CSV containing:
    \begin{itemize}[label=--]
        \item Code\_Name
        \item Definition
    \end{itemize}
    \item A Responses CSV where each row is a participant response. It contains:
    \begin{itemize}[leftmargin=*,label=--]
        \item Response\_ID
        \item Response\_Text
        \item Codes (as a comma-separated list of strings)
    \end{itemize}
\end{enumerate}

\vspace{\baselineskip}
\noindent
Your task:\par\noindent
Assign one or more codes to the responses based on the codebook definitions. Two remarks:
\begin{itemize}[leftmargin=*,label=--]
    \item The first 100 responses were already labeled with the codes in the codebook. You will assign codes to the remaining responses.
    \item Each code explains why the patch was selected, not why it was not selected. So, the perspective is always on the selected patch. If the response focuses only on the not-selected patch, the code must be about the selected patch.
\end{itemize}

\vspace{\baselineskip}
\noindent
Rules:\par\noindent
\begin{itemize}[leftmargin=*,label=--]
    \item Assign multiple codes if appropriate.
    \item Prioritize existing codes in the codebook.
    \item If no existing code adequately fits, you may create and use a new code, but only if absolutely necessary.
    \begin{itemize}[leftmargin=*,label=--]
        \item Any new code must have:
        \begin{itemize}[leftmargin=*,label=--]
            \item The code name
            \item A concise definition consistent with the existing codebook style
        \end{itemize}
    \end{itemize}
    \item Keep coding consistent across similar responses.
    \item Do not modify any response data or existing codes assigned to the first 100.
\end{itemize}

\vspace{\baselineskip}
\noindent
Output requirements:\par\noindent
Return the Responses CSV with only the responses you labeled and the added codes under ``Codes''. Keep the same format as the original CSV file.

\vspace{\baselineskip}
\noindent
If any new codes are created, append:\par\noindent
NEW CODES CREATED:\par\noindent
Code\_Name $\vert$ Definition

\vspace{\baselineskip}
\noindent
If you had problems with assigning codes to certain entries, append their IDs and a brief explanation of the problem you had.

\vspace{\baselineskip}
\noindent
Input:\par\noindent
CODEBOOK CSV: [ATTACHED FILE]\par\noindent
RESPONSES CSV: [ATTACHED FILE]

\vspace{\baselineskip}
\noindent
Output only the updated CSV, the new codes (if any), and the problem you had (if any). No additional commentary.
}

\section{\deepseekShort's Prompt for the Code Assignment (\rqq{4})}
\label{appendix:prompt_rq4}

The following prompt was given to \deepseekLong through the \textsc{ChatAI} service of \textsc{Academic Cloud} (\url{https://chat-ai.academiccloud.de/}) on October 1, 2026.
This service hosts several models, including \deepseekLong, in Germany and guarantees that the data is processed in loco, securely stored, and not used for training other models.
The model was queried with a \textit{temperature} of 0.5, \textit{top\_p} of 0.5, and a ``low'' reasoning setting.

\prompt{}{
You are an expert in thematic analysis and in software security. Your goal is to classify a set of participant responses and perform qualitative open coding. These responses come from a questionnaire administered during an experiment in which participants were asked to select one of two patches that fixed a certain vulnerability (the one believed to be correct). Then, once they were informed about the provenance of the two patches (one came from a human and one from an LLM), they could optionally change their selection and motivate. The responses you will review are the motivations they disclosed to explain their second decision.

\vspace{\baselineskip}
\noindent
You will be given:
\begin{enumerate}[leftmargin=*,label=\arabic*.]
    \item A Responses CSV where each row is a participant response. It contains:
    \begin{itemize}[label=--]
        \item Response\_ID
        \item 1st\_Response\_Text
        \item 2nd\_Response\_Text
        \item Provenance\_ACK (yes/no)
        \item NO\_ACK\_Class (string)
        \item YES\_ACK\_Codes (as a comma-separated list of strings)
    \end{itemize}
    \item A Codebook CSV containing the candidate codes to use for YES\_ACK\_Codes. It contains:
    \begin{itemize}[leftmargin=*,label=--]
        \item Code\_Name
        \item Definition
    \end{itemize}
\end{enumerate}

\vspace{\baselineskip}
\noindent
Your task:\par\noindent
Review the second responses (`2nd\_Response\_Text`) and determine if they acknowledged the provenance of the patches, assigning ``yes'' or ``no''. You can use the first response (`1st\_Response\_Text`) for context.
Then, for the ``no'' cases, classify `NO\_ACK\_Class` into one of these possibilities:

\begin{itemize}[leftmargin=*,label=--]
    \item ``Restate original motivations'', if the second response merely states the same motivations given already in the first response.
    \item ``Provide additional motivations'', if the second response provides some more motivations than those given in the first response.
    \item ``Unclear/Unspecified motivations'', in all other cases.
\end{itemize}

For the "yes" cases, assign one or more codes based on the codebook definitions.

Remark:
\begin{itemize}[leftmargin=*,label=--]
    \item The first 100 responses have already been analyzed. Use these as guidance. You will work on the remaining responses.
\end{itemize}

\vspace{\baselineskip}
\noindent
Rules:\par\noindent
\begin{itemize}[leftmargin=*,label=--]
    \item Assign multiple codes to `YES\_ACK\_Codes` if appropriate.
    \item Prioritize existing codes in the codebook.
    \item If no existing code adequately fits, you may create and use a new code, but only if absolutely necessary. 
    \begin{itemize}[leftmargin=*,label=--]
        \item Any new code must have:
        \begin{itemize}[leftmargin=*,label=--]
            \item The code name
            \item A concise definition consistent with the existing codebook style
        \end{itemize}
    \end{itemize}
    \item Keep coding consistent across similar responses.
    \item Do not modify any response data or existing codes assigned to the first 100.
\end{itemize}

\vspace{\baselineskip}
\noindent
Output requirements:\par\noindent
Return the Responses CSV with only the responses you labeled and the added labels under `Provenance\_ACK`, `NO\_ACK\_Class`, and `YES\_ACK\_Codes`. Keep the same format as the original CSV file.

\vspace{\baselineskip}
\noindent
If any new codes are created, append:\par\noindent
NEW CODES CREATED:\par\noindent
Code\_Name $\vert$ Definition

\vspace{\baselineskip}
\noindent
If you had problems with assigning codes to certain entries, append their IDs and a brief explanation of the problem you had.

\vspace{\baselineskip}
\noindent
Input:\par\noindent
RESPONSES CSV: [ATTACHED FILE]\par\noindent
CODEBOOK CSV: [ATTACHED FILE]

\vspace{\baselineskip}
\noindent
Output only the updated CSV, the new codes (if any), and the problem you had (if any). No additional commentary.
}

\end{document}